\documentclass[usenatbib]{mnras}

\usepackage[T1]{fontenc}
\usepackage{ae,aecompl}

\usepackage{graphicx}
\usepackage{amsmath}
\usepackage{amssymb}

\renewcommand{\vec}[1]{\bmath{#1}}
\newcommand{\transpose}[1]{{#1}^{\mathrm{T}}}
\newcommand{\mat}[1]{\bmath{\mathsf{#1}}}
\newcommand{\matt}[1]{\transpose{\mat{#1}}}
\newcommand{\bigo}[1]{\mathcal{O}\left(#1\right)}
\newcommand{\abs}[1]{\left|#1\right|}

\newcommand\ionl[3]{\ion{#1}{#2}$\;\lambda${#3}\relax}
\newcommand{\mtxt}[2]{{#1}_{\mathrm{#2}}}
\newcommand{\ltxt}[1]{\mtxt{\lambda}{#1}}
\newcommand{\lobs}{\ltxt{obs}}
\newcommand{\lrest}{\ltxt{rest}}
\newcommand{\covar}[2]{\sigma_{#1,#2}^2}
\newcommand{\ccf}{\mathrm{ccf}}
\newcommand{\erf}{\mathrm{erf}}
\newcommand{\transition}[3]{\mathrm{tr}_{#1}^{#2}(#3)}

\title[QSO characterization within the Gaia mission]{Determination of astrophysical parameters of quasars within the \textit{Gaia} mission}

\author[L. Delchambre]{L. Delchambre$^{1}$\thanks{E-mail:	ldelchambre@ulg.ac.be} \\
$^{1}$ Institut d'Astrophysique et de G\'eophysique, Universit\'e de Li\`ege, All\'ee du 6 Ao\^ut 17, B-4000 Sart Tilman (Li\`ege), Belgique}

\date{Accepted ???. Received ???; in original form ???}

\pagerange{\pageref{firstpage}--\pageref{lastpage}}
\pubyear{2017}

\begin{document}

\maketitle

\label{firstpage}

\begin{abstract}
We describe methods designed to determine the astrophysical parameters of quasars based on spectra coming from the red and blue spectrophotometers of the \textit{Gaia} satellite. These methods principally rely on two already published algorithms that are the weighted principal component analysis and the weighted phase correlation. The presented approach benefits from a fast implementation; an intuitive interpretation as well as strong diagnostic tools on the potential errors that may arise during predictions. The production of a semi-empirical library of spectra as they will be observed by \textit{Gaia} is also covered and subsequently used for validation purpose. We detail the pre-processing that is necessary in order for these spectra to be fully exploitable by our algorithms along with the procedures that are used in order to predict the redshifts of the quasars; their continuum slopes; the total equivalent width of their emission lines and whether these are broad absorption line (BAL) quasars or not. Performances of these procedures were assessed in comparison with the Extremely Randomized Trees learning method and were proven to provide better results on the redshift predictions and on the ratio of correctly classified observations though the probability of detection of BAL quasars remains restricted by the low resolution of these spectra as well as by their limited signal-to-noise ratio. Finally, the triggering of some warning flags allows us to obtain an extremely pure subset of redshift predictions where approximately $99\%$ of the observations come along with absolute errors that are below $0.1$.
\end{abstract}


\section{Introduction}
\label{sec:intro}

	\textit{Gaia} is one of the cornerstone space missions of the Horizon 2000+ science program of the ESA that aims to bring a consensus on the history and evolution of our Galaxy through the survey of a billion celestial objects \citep{perryman2001}. This objective being achieved by capturing a `snapshot' of the present structure; dynamic and composition of the Milky Way by means of precise astrometric and photometric measurements of all the observed objects as well as by the determination of the distances; proper motions; radial velocities and chemical compositions of a subset of these objects \citep{gaia2016}.
	
	The on-board instrumentation is principally composed of two $1.45 \times 0.50$ m telescopes pointing in directions separated by a basic angle of $106.5\degr$, the light acquisition being then carried out by slowly rotating the satellite on its spin axis while reading each CCD column at the same rate as the objects cross the focal plane (i.e. in the so-called Time Delay Integration mode, hereafter TDI mode). The high astrometric precision of \textit{Gaia} then coming from: (i) its lack of atmospheric perturbations (ii) its large focal length of $35$ m (iii) the combination of the beams of light coming from both telescopes onto a single focal plane composed of a patchwork of $106$ CCDs that allow to relate the positions of the objects coming from the two fields of view with an extremely precise angular resolution (iv) its scanning law that enables to maximize the number of observed objects as well as the number of positional relations arising from the previous point \citep{lindegren2012}.

	In addition, a high-resolution ($R = \lambda/\Delta \lambda = 11,700$) spectrometer, called Radial Velocity Spectrometer (RVS), centred around the $\ion{Ca}{II}$ triplet ($845$-$872$ nm) will allow to determine the radial velocities of some of the most luminous stars ($G_{\text{RVS}} < 16$ mag), while two low-resolution spectrophotometers, namely the Blue Photometer (BP) observing in the range $330$-$680$ nm ($13 < R < 85$) and the Red Photometer (RP) observing in the range $640$-$1050$ nm ($17 < R < 26$), will allow to classify and characterize the objects having $G < 20$ mag. The interested reader is invited to read \citet{debruijne2012} and \citet{gaia2016} for a more complete description of the \textit{Gaia} spacecraft and of its payload.
	
	The previously described instrumentation, combined with the fact that \textit{Gaia} is a full-sky survey where each object will be observed $70$ times on average, is a unique opportunity in order to achieve some additional objectives. A non-exhaustive list of such applications are: a finer calibration of the whole cosmological distance ladder (i.e. through parallaxes, Cepheids \& RR Lyrae stars, type-Ia supernovae, ...) \citep{debruijne2012}; a better understanding of the stellar physics and evolution through the refinement of the Hertzsprung-Russell diagram \citep{jordi2008}; the discovery of thousands high-mass exoplanets \citep{perryman2014} as well as new probes regarding fundamental physics \citep{mignard2009}.
	
	Amongst the most peculiar objects that \textit{Gaia} will observe, stand quasars--also termed quasi-stellar objects (QSOs) for historical reasons. Quasars are active galactic nuclei originating from the matter accretion that was occurring in the vicinity of supermassive black holes being at cosmological distances. Due to their high luminosity ($L > 10^{12} L_{\sun}$) and their large redshift ($0 < z < 7$), these objects play a key-role in fixing the celestial reference frame used by \textit{Gaia}, but they also have their own intrinsic interest in various cosmological applications like in the evolution scenarios of the galaxies \citep{hamann1999}; as discriminants over the various universe model and their parametrization \citep{lopezcorredoira2016}; as tracers of the large-scale distribution of Baryonic matter at high redshift \citep{yahata2005} or as a means to independently constrain the Hubble constant if the latter are gravitationally lensed \citep{schneider2013}.
	
	The identification and characterization of the $500,000$ quasars that \textit{Gaia} is expected to observe takes place in the framework of the Data Processing and Analysis Consortium (DPAC) which is responsible for the treatment of the \textit{Gaia} data in a broad sense, that is: from data calibration and simulation to catalogue publication through intermediate photometric/astrometric/spectroscopic processing; variability analysis and astrophysical parameters (APs) determination. The DPAC is an academic consortium composed of nine Coordination Units (CUs), each being in charge of a specific part of the data processing \citep{omullane2007}. One of these, the CU8 `Astrophysical Parameters', is dedicated to the classification of the objects observed by \textit{Gaia} and to the subsequent determination of their APs \citep{bailerjones2013}.
	
	The present paper describes the algorithms that are to be implemented within the CU8 Quasar Classifier (QSOC) software module in order to determine the APs of the objects classified as QSOs by the CU8 Discrete Source Classifier (DSC) module while relying on their low-resolution BP/RP spectra. The collected APs aiming to be published within the upcoming Gaia data release 3 catalogue. The covered APs encompass: the redshift; the QSO type (i.e. type I/II QSO or Broad Absorption Line QSO, hereafter BAL QSO); the slope of the QSO continuum and the total equivalent width of the emission lines.
	
	In the following, section \ref{sec:conventions} explains the conventions we used along this paper. Section \ref{sec:methods} makes a brief review of the methods that were specifically developed in the field of this study. We present the production of a semi-empirical library of BP/RP spectra of QSOs used in order to train/test our models in section \ref{sec:bprp_library_building}. The AP determination procedures are covered within section \ref{sec:ap_determination} while their performances are assessed in section \ref{sec:comparison}. Some discussion on the latter takes place in section \ref{sec:discussion}. Finally, we conclude in section \ref{sec:conclusion}.
	
\section{Conventions}
\label{sec:conventions}

	This paper uses the following notations: vectors are in bold italic, $\vec{x}$; $x_i$ being the element $i$ of the vector $\vec{x}$. Matrices are in uppercase boldface or are explicitly stated; i.e. $\mat{X}$ from which element at row (variable) $i$, column (observation) $j$ will be denoted by $\mat{X}_{ij}$.

	\textit{Flux} will here denote the spectral power received per unit area (derived unit of W m$^{-2}$) while \textit{flux density} will represent the received flux per wavelength unit (derived units of W m$^{-3}$). If not stated otherwise, input spectral energy distributions (SED) will be considered to be expressed in terms of flux density while BP/RP spectra will be expressed in terms of flux by convention.

\section{Methods}
\label{sec:methods}

	One of the main characteristics of the \textit{Gaia} data processing is the large amount of information that has to be handled (e.g. about $40$ gigabytes of compressed scientific data are received from the satellite each day) and the consequent need for fast and reliable algorithms in order to reduce those data. These last requirements led the CU8 scientists to use techniques coming nearly exclusively from the field of the supervised learning methods whose underlying principle is to guess the APs of each observed object, which are unknowns, based on the interpolation of the APs of some similar template objects \citep{bailerjones2013}.

	These methods have been proven to be fairly fast and reliable but often consist in black-box algorithms having no physical significance and having only basic diagnostic tools in order to identify the potential problems that may occur during the APs retrieval. This last point is particularly crucial in the case of medium-to-low quality observations, like BP/RP spectra, or in the case where the problem is itself prone to error, like the existing degeneracy in the redshift determination of low signal-to-noise ratio (SNR) QSOs \citep{delchambre2016}.
	
	With these constraints and shortcomings in mind, we have developed two complementary algorithms that are specifically designed to gather the quasar APs within the \textit{Gaia} mission based on the object BP/RP spectra while providing a clear diagnostic tool and ensuring an execution time that is limited to $\bigo{N \log N}$ floating point operations (i.e. conventionally considered as `fast' algorithms).

\subsection{Weighted principal component analysis}
\label{subsec:wpca}

	Principal component analysis (PCA) aims to extract a set of templates -- the principal components -- from a set of observations while retaining most of its variance \citep{pearson1901}. Mathematically, it is equivalent to find a decomposition of the covariance matrix associated with the input data set,
\begin{equation}
\label{eq:pca_eigen}
\mat{\sigma^2} = \mat{P} \mat{D} \matt{P},
\end{equation}
that is such that $\mat{P}$ is orthogonal and $\mat{D}$ diagonal and for which
\begin{equation}
\label{eq:pca_di_dj}
\mat{D}_i \geq \mat{D}_j; \; \forall i < j.
\end{equation}
The first columns of $\mat{P}$ being then the searched principal components. A decomposition such as the one of equation \ref{eq:pca_eigen} is straightforwardly given by the singular value decomposition (SVD) of the covariance matrix \citep{press2002}.

	Consider now the building of a set of rest-frame quasar templates based on a spectral library having a finite precision on the fluxes and a limited wavelength coverage. From the definition of the redshift, we will have that the observed wavelength, $\lobs$, can be related to the rest-frame wavelength, $\lrest$, through
\begin{equation}
\lobs = ( z + 1 ) \lrest,
\end{equation}
and as a consequence, we will have that every quasar within the input library will cover a specific rest-frame wavelength range that depends on its redshift. Furthermore, the measurement of the quasar fluxes often comes along with an estimation of their associated uncertainties originating, for example, from the Poisson nature of the photons counting; from the CCD readout noise; from the sky background subtraction or from spectra edge effects. These uncertainties being not taken into account within the classical PCA implementation.

	In \citet{delchambre2015}, we solved the previously mentioned issues by considering the use of a weighted covariance matrix inside equation \ref{eq:pca_eigen}. For this purpose, we defined the weighted covariance of two discrete variables, $\vec{x}$ and $\vec{y}$ having weights respectively given by $\vec{w^x}$ and $\vec{w^y}$ and weighted mean values given by $\bar{x}$ and $\bar{y}$ as
\begin{equation}
\covar{\vec{x}}{\vec{y}} = \dfrac{\sum_i \left(x_i - \bar{x} \right) w_i^x w_i^y \left( y_i - \bar{y} \right)}{\sum_i w_i^x w_i^y}.
\label{eq:pca_covar}
\end{equation}

	The suggested implementation relies on two spectral decomposition methods, namely the power iteration method and the Rayleigh quotient iteration, that allow to gain flexibility; numerical stability as well as lower execution times\footnote{Under the condition that the number of observations within the input data set is much larger than the number of variables} when compared to alternative weighted PCA methods \citep{bailey2012,tsalmantza2012}.

\subsection{Weighted phase correlation}
\label{subsec:wpcorr}

	The redshift has a particular importance over the whole quasar APs because any error committed on the latter would make the other APs diverge. Its is then critical to have the most precise estimation of it along with a strong diagnostic tool in order to flag the insecure predictions. A technique fulfilling these requirements stands in the phase correlation algorithm \citep{glazebrook1998} whose goal is to find the phase at which an input signal and a set of templates match at best in a chi-squared sense. 
	
	For reasons already enumerated within section \ref{subsec:wpca}, we will consider here a weighted version of the previously mentioned algorithm. We are then searching for the shift at which an input spectrum, $\vec{s}$, associated with a weight vector, $\vec{w}$, and a set of templates, $\mat{T}$, matches at best in a weighted chi-squared sense. Mathematically, it is equivalent to find the shift, $Z$, for which
\begin{equation}
\label{eq:wpcorr_chi2z}
\chi^2(Z) = \sum_i w_i^2 \left( s_i - \sum_j a_j(Z) \mat{T}_{(i+Z) j} \right)^2
\end{equation}
is minimal given that $\vec{a}(Z)$ are the linear coefficients minimizing equation \ref{eq:wpcorr_chi2z} for a specific shift.
	
	In \citet{delchambre2016}, we showed that the latter equation can be re-written as
\begin{equation}
\label{eq:wpcorr_chi2z_bis}
\chi^2(Z) = \sum_i w_i^2 s_i^2 - \ccf(Z),
\end{equation}
where $\ccf(Z)$ is the so-called \textit{cross-correlation function} (CCF) at shift $Z$ which can be evaluated for all $Z$ in $\bigo{N \log N}$ floating point operations, $N$ being the number of samples we used. Given that the first term of equation \ref{eq:wpcorr_chi2z_bis} is independent of the shift, we will simply have that the minimum of equation \ref{eq:wpcorr_chi2z} will be associated with the maximum of the CCF.

	Practically, both $\vec{s}$ and $\mat{T}$ must be sampled on a uniform logarithmic wavelength scale in order for the redshift to turn into a simple linear shift (i.e. $\log \lobs = \log(z+1) + \log \lrest$) and must be extended and zero-padded such as to deal with the periodic nature of the phase correlation algorithm. A sub-sampling precision on the shift can then be gained by fitting a quadratic curve in the vicinity of the optimum of the CCF while the curvature of this quadratic curve will be used as an approximation of the uncertainty associated with the found shift.
	
	The described weighted phase correlation algorithm relies on the assumption that the most probable redshift is associated with the maximal peak of the CCF, which is not always verified in the case of QSOs. The reason for this is twofold:
\begin{enumerate}
\item The highest peak of the CCF may not always lead to a physical solution like the omission of some characteristic emission lines (e.g. Ly$\alpha\;\lambda 121$; \ionl{Mg}{ii}{279}; or H$\alpha\;\lambda 656$ nm) or the fit of a `negative' emission line coming from the presence of matter being in the line-of-sight towards the observed QSO. The origin of this issue mainly stands in the imperfections of the templates we used as well as in the assumption we made that quasar spectra can be modelled as a linear combination of these templates.
\item In the case of low-SNR spectra, it may also happen that the signal of some emission lines starts to be flooded within the noise such that these will not be recognized as a genuine signal but rather as variance coming from noise. As a result, ambiguities can emerge within the CCF (i.e. multiple equivalent maxima) and hence within the redshift determination.
\end{enumerate}

\begin{table}
\caption{Rest-frame wavelengths and relative intensities of the emission lines used in the computation of $\mtxt{Z}{score}(z)$. Listed values come from the PCA mean spectra described in section \ref{sec:ap_determination}.}
\label{tbl:zscore_params}
\begin{tabular}{ccc}
\hline
Emission & Rest-frame wavelength & Relative intensity \\
line(s) & nm & compared to Ly$\alpha$ \\
\hline
\ion{O}{vi} & $103.07$ & $0.13161$ \\ 
Ly$\alpha$ & $121.81$ & $1.00000$ \\ 
\ion{C}{iv} & $154.63$ & $0.30834$ \\
\ion{C}{iii]} & $190.24$ & $0.16413$ \\
\ion{Mg}{ii} & $280.18$ & $0.21778$ \\
H$\beta$+$[$\ion{O}{iii}$]$ & $488.06$ & $0.42182$ \\
H$\alpha$ & $ 656.86$ & $1.17143$ \\
\hline
\end{tabular}
\end{table}

	In order to identify these sources of errors, we defined two complementary score measures associated with each redshift candidate: (i) $\chi_r^2(z)$, defined as the ratio of the value of the CCF evaluated at $z$ to the maximum of the CCF and (ii) $\mtxt{Z}{score}(z)$ defined as
\begin{equation}
\label{eq:wpcorr_zscore}
\mtxt{Z}{score}(z) =  \prod_\lambda \left[ \frac{1}{2} \left(1 + \erf \frac{e_\lambda}{\sigma(e_\lambda) \sqrt{2}}\right) \right]^{I_\lambda},
\end{equation}
where $e_\lambda$ is the mean value of the continuum-subtracted emission line standing at rest-frame wavelength $\lambda$ if we consider the observed spectrum to be at redshift $z$; $\sigma(e_\lambda)$ is the associated uncertainty and $I_\lambda$ is the theoretical intensity of the emission line standing at $\lambda$ normalized such that all the covered emission lines intensities sum up to one. Equation \ref{eq:wpcorr_zscore} can then be seen as the weighted geometric mean of a set of normal cumulative distribution functions of mean zero and standard deviations $\sigma(e_\lambda)$ evaluated in $e_\lambda$. Table \ref{tbl:zscore_params} summarizes the various emission lines and theoretical intensities we used in the context of the present study.

	We can already notice that these two score measures can easily highlight the sources of errors that may occur within the CCF peak selection, namely the choice of an unphysical solution and the choice of an ambiguous candidate respectively through a low $\mtxt{Z}{score}$ and through a low absolute difference between the chosen candidate's $\chi_r^2$ and the one from another candidate. These will constitute \textit{in fine} strong diagnostic tools regarding our implementation.

\section{Semi-empirical BP/RP spectral library building}
\label{sec:bprp_library_building}

	Similarly to supervised learning methods, the undertaken approach is based on the availability of a learning library of BP/RP spectra for which the various APs are known. Such a library being non-existent at the present time, we had to convert an already released spectral library of QSOs into BP/RP spectra according to the most up-to-date instrument model. We focused, for this purpose, on the twelfth data release of the Sloan Digital Sky Survey quasar catalogue \citep[hereafter DR12Q]{paris2017}. The choice of this specific catalogue comes from: (i) the fact that each spectrum in it was visually inspected yielding extremely secure APs (ii) the large number of $297,301$ QSOs it contains (amongst which $29,580$ BAL QSOs) (iii) its medium resolution of $1300 < R < 2500$ (iv) its spectral coverage which is comparable to the one of the \textit{Gaia} BP/RP spectrophotometers ($360 < \lambda < 1000$ nm).
	
	This spectral library will then have to be extended such as to match the wavelength range covered by the Gaia BP/RP spectra and be subsequently convolved with the instrumental response of the BP/RP spectrophotometers such as to provide the final library.

\subsection{Spectra extrapolation}
\label{subsec:spectra_extrapolation}

	Besides the fact that the DR12Q spectra have a narrower wavelength coverage ($360 < \lambda < 1000$ nm) when compared to the BP/RP spectra ($330 < \lambda < 1050$ nm), we also have to mention that the regions where $\lambda < 380$ and where $\lambda > 925$ nm often tend to be unreliable because of spectrograph edge effects while some other inner regions might be discarded because of bad CCD columns, cosmic rays, significant scattered light or sky background subtraction problems, for example \citep{bolton2012,dawson2013}. In order to solve these shortcomings, we have extracted a set of rest-frame PCA templates out of the DR12Q library that were later fitted to each individual spectrum as a mean to extrapolate them. Note that since the DR12Q spectra are already sampled on a logarithmic wavelength scale (at a sampling rate of $\Delta \log_{10} \lambda = 10^{-4}$), no re-sampling will be needed before extrapolation.
	
\begin{figure*}
\includegraphics{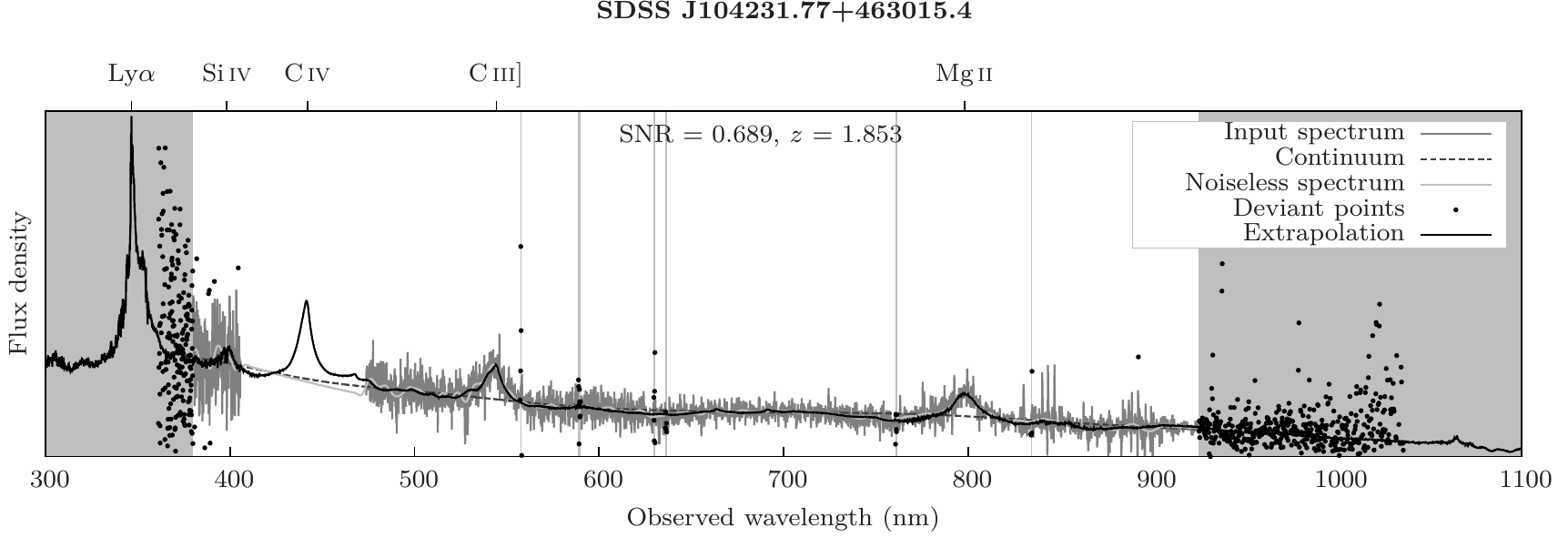}
\caption{Results of the pre-processing and extrapolation procedures. The input spectrum is decomposed into a continuum part, a noiseless spectrum and a set of deviant points (night sky emission lines and spectrograph edge effects being here depicted as shaded regions). This decomposition allows us to compute the SNR associated with each spectrum and to set those having a SNR $> 1$ on a common logarithmic rest-frame wavelength scale such as to extract PCA templates out of them. The fit of these templates to each input spectrum provides the final extrapolation. Note that the illustrated input spectrum has some corrupted samples in the region encompassing the \ion{C}{iv} emission line which are successfully recovered through this extrapolation procedure. Similarly, the unobserved Ly$\alpha$ emission line seems to be consistently reproduced as well.}
\label{fig:extrapolation}
\end{figure*}
		
	Raw spectra are not readily exploitable, rather they have first to be pre-processed such as to get rid of contaminating signals and to have some insights about their usability. For this purpose, we used a procedure that is identical to the one described in \citet[section 5.1]{delchambre2016}. We will hence concentrate here on the results of this procedure rather than on the underlying implementation details. We will then get, for each spectrum: (i) the set of deviant points coming from a $k$--$\sigma$ clipping algorithm applied to the high frequency components of the spectrum as well as from the removal of the night sky emission lines and spectrograph edge effects (ii) an empirical estimation of the QSO continuum coming from the low frequency components of a multi-resolution analysis of the spectrum (iii) a smoothed version of the provided spectrum, that we will consider here as being noiseless (iv) an evaluation of its SNR coming from the ratio of the variance that is present within the noiseless and continuum-subtracted spectrum over the variance that can be attributed to noise (i.e. raw fluxes from which we subtracted the deviant points; the QSO continuum and the noiseless spectrum). Figure \ref{fig:extrapolation} illustrates the results of the previously described procedure and provides a self-explanatory example of the necessity we have to pre-process our input spectra.
	
\begin{figure*}
\includegraphics{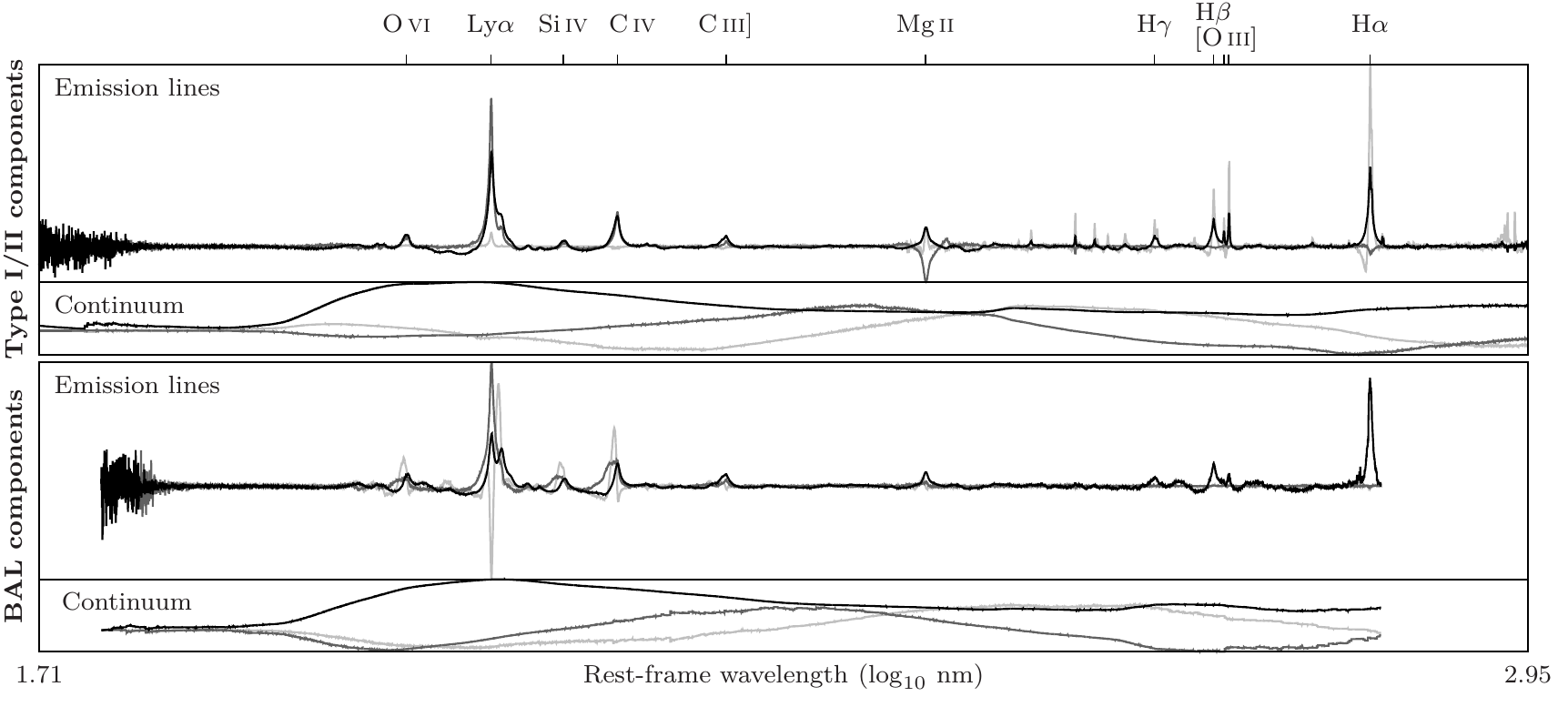}
\caption{Mean observations (black lines) and first two principal components coming from the PCA decompositions of the DR12Q emission lines and continuum spectra regarding type I/II QSOs (up) and BAL QSOs (down). Notice how the entire emission lines are modelled by the type I/II QSOs templates while the BAL QSOs templates rather focus on the \ion{O}{vi}; Ly$\alpha$; \ion{Si}{iv} and \ion{C}{iv} emission lines where most of the BAL characteristics are found.}
\label{fig:pca_extrapolation}
\end{figure*}

	Spectra having a SNR larger than one are then normalized such as to have a weighted norm of one and are subsequently set on a common logarithmic rest-frame wavelength scale. These will constitute the input dataset upon which we will extract our PCA templates. We choose to consider the retrieval of the BAL QSO templates  separately from the type I/II QSOs as a way to ensure that the characteristic features of the BAL QSOs are correctly reproduced within our extrapolated spectra. Doing otherwise would have required a much larger number of PCA components to be fitted in order to accurately model those features (the latter being not seen in the vast majority of QSOs, they would have been omitted from the dominant PCA components). Also, we required the continuum to rely on an empirical basis such as to restrain at most any unphysical behaviour within our extrapolation. We will consequently subtract each empirical continuum from each input spectrum as a way to extract the PCA components from both these subtracted spectra as well as from the continua themselves. By way of comparison, continuum templates are frequently taken as being a combination of power-law and exponential functions \citep{claeskens2006} that often succeed in reproducing the observed spectrum but that tend to diverge over the unobserved wavelengths. Consequently, four sets of PCA templates were built based upon the algorithm described in section \ref{subsec:wpca}: one set of templates for the type I/II QSO emissions lines; one similar set associated with the BAL QSOs and two corresponding sets of continuum templates. Figure \ref{fig:pca_extrapolation} provides the mean observation and two first principal components for these four sets of templates. For the sake of completeness, let us also mention that, during PCA retrieval, weights were taken as the inverse standard deviation on the fluxes regarding the emission line PCAs while these were simply set to one if the continuum we computed was associated with some observed fluxes and zero otherwise.

	Finally, we used $15$ emission line templates (the mean observation and the first $14$ PCA components) for both the fit of the type I/II and BAL observations along with $5$ continuum templates for the type I/II QSOs and $6$ continuum templates for the BAL QSOs. These fits ultimately provide the extrapolated spectra, as illustrated in figure \ref{fig:extrapolation}. The number of templates we used allows us to explain $68.27$\% of the weighted variance\footnote{Weighted variance naturally results from equation \ref{eq:pca_covar} in the case where $\vec{x} = \vec{y}$ while having expected values, $\bar{x} = 0$ in the case of the weighted variance of the input data set or equal to the optimal linear combination of the templates in the case of the explained weighted variance.} that is present within the type I/II emission lines and $99.82$\% of their continuum variance. These become respectively $66.1$\% and $99.85$\% in the case of BAL QSOs. Even if apparently low, these ratios practically reflect the fact that the spectra we used come along with noise that will not be grabbed by the dominant PCA components. Consequently, the produced extrapolation will be considered here as being noiseless.
	
\subsection{BP/RP instrumental convolution}
\label{subsec:bprp_instrumental_convolution}

	The optical system of the \textit{Gaia} BP/RP spectrophotometers consists for each in six mirrors, a dispersing prism and a set of dedicated CCDs (either blue or red-enhanced) that can be modelled as
\begin{equation}
\label{eq:bprp_model}
S(x) = \int N_\lambda R_\lambda L_\lambda(x - x_\lambda)  d\lambda,
\end{equation}
where $S(x)$ is the dispersed flux in the one-dimensional co-moving\footnote{sliding at the same rate as the TDI mode drift each CCD column} focal plane position $x$; $N_\lambda$ is the input SED at the observed wavelength $\lambda$; $R_\lambda$ is the global instrumental response at $\lambda$ and where $L_\lambda(x-x_\lambda)$ is the monochromatic line spread function (LSF) standing at $\lambda$ and being evaluated at $x-x_\lambda$, $x_\lambda$ being the co-moving focal plane position associated with $\lambda$. In more details, the global instrumental response $R_\lambda$ encompasses: the mirrors reflectivity, the attenuation that is due to particulate and molecular contamination, the attenuation coming from the mirror roughness, the prism transmissivity and the CCD quantum efficiency at the observed wavelength $\lambda$. Note that in the following, we will consider a mean instrumental model averaged over each field-of-view and over each row of CCD within the focal plane given that our goal is to simulate end-of-mission (combined) spectra that will consist in an aggregation of individual (epoch) observations.

	The BP/RP sampled fluxes, $\vec{s}$, will then be given by
\begin{equation}
\label{eq:bprp_model_sampled}
s_i = \int_{-0.5}^{+0.5} S\left(x + \frac{i}{\mtxt{N}{over}}\right) dx,
\end{equation}
where $\mtxt{N}{over}$ is the oversampling we choose to use. This oversampling arise from the higher SNR that is gained by the combination of the epoch spectra which, in turn, provide the opportunity to reach a higher sampling rate when compared to the initial $60$ pixels provided by the BP/RP acquisition window (e.g. through flux interpolation). A common consensus within the CU8 is to consider an oversampling of $\mtxt{N}{over} = 8$, which results in $480$ samples for each of the BP and RP spectra. This convention will be adopted here.

	The instrument model described so far is not able to deal with the various sources of noise that will contaminate our actual observations. Instead, the spectra produced through equations \ref{eq:bprp_model} and \ref{eq:bprp_model_sampled} will consist in the approximated noise-free counterparts of what \textit{Gaia} will observe. Extending the aperture photometry approach developed in \citet{jordi2010}, we can still have an estimation of the noise variance that is associated with each sampled flux, $s_i$, as
\begin{equation}
\label{eq:noise_model_general}
\sigma_i^2 = m^2 \frac{\mtxt{\sigma}{epoch}^2}{\mtxt{N}{epoch} / \mtxt{N}{over}} + \mtxt{\sigma}{cal}^2,
\end{equation}
where $m$ is an overall mission safety margin designed to take into account the potentially unknown sources of errors ($m = 1.2$, by convention); $\mtxt{\sigma}{epoch}^2$ is the variance of the noise associated with $s_i$ if the latter was coming from a single epoch observation; $\mtxt{N}{epoch}$ is the number of epoch observations used to compute the combined spectra and $\mtxt{\sigma}{cal}^2$ is the uncertainty arising from the flux internal calibration. The scaling of the single epoch variance, $\mtxt{\sigma}{epoch}^2$, reflects the assumption we made that each flux within the combined spectra comes from the mean value of a set of $\mtxt{N}{epoch} / \mtxt{N}{over}$ epoch fluxes. In extreme cases, for example, we will have that each combined flux is averaged over the whole epoch observations (i.e. in the case of $\mtxt{N}{over} = 1$) while in the case of $\mtxt{N}{over} = \mtxt{N}{epoch}$, each flux within the combined spectra can be seen as gathered directly from the epoch spectra. Next, the variance coming from the uncertainties in the flux internal calibration, $\mtxt{\sigma}{cal}^2$, is taken to be equal to the inner product of the fluxes that are present within the pixels surrounding each sample with a linear function that is inversely proportional to the global instrumental response, $R_\lambda$, evaluated in those pixels. Its objective being to take into account the fact that the precision on the flux calibration will principally depend on the instrumental response in the vicinity of the pixel of interest. Because of the intricacy that is inherent to the modelling of these calibration errors, the latter were voluntarily tuned such as to stand within a moderate range of values.

	We can then decompose the epoch variance, $\mtxt{\sigma}{epoch}^2$ into variance coming from photon and CCD noise, $\mtxt{\sigma}{flux}^2$, and variance coming from the uncertainties in the background estimation, $\mtxt{\sigma}{bg}^2$, as
\begin{equation}
\label{eq:noise_epoch}
\mtxt{\sigma}{epoch}^2 = \frac{\mtxt{\sigma}{flux}^2 + \mtxt{\sigma}{bg}^2}{\tau^2},
\end{equation}
where $\tau$ is the effective CCD exposure time, the latter being dependent on the $G$ magnitude of the objects based on the activation of bypasses within some specific CCD columns which aim to prevent luminous objects from saturating \citep{debruijne2012}. Both terms within equation \ref{eq:noise_epoch} can then be simply extended as
\begin{eqnarray}
\label{eq:noise_flux}
\mtxt{\sigma}{flux}^2 & = & (s_i + b) \tau + r^2, \\
\label{eq:noise_bg}
\mtxt{\sigma}{bg}^2 & = & \frac{b \tau + r^2}{\mtxt{N}{bg}},
\end{eqnarray}
where $r$ is the total CCD detection noise including, amongst other, the CCD readout noise and the CCD dark noise; $b$ is the background flux we subtracted from our observation and that we will consider here as a constant based on a typical sky-background surface brightness and where $\mtxt{N}{bg}$ is the number of pixels we used in order to estimate $b$, that is taken here as being equal to the width of the BP/RP acquisition window (i.e. $\mtxt{N}{bg} = 12$ pixels).

\begin{figure*}
\includegraphics{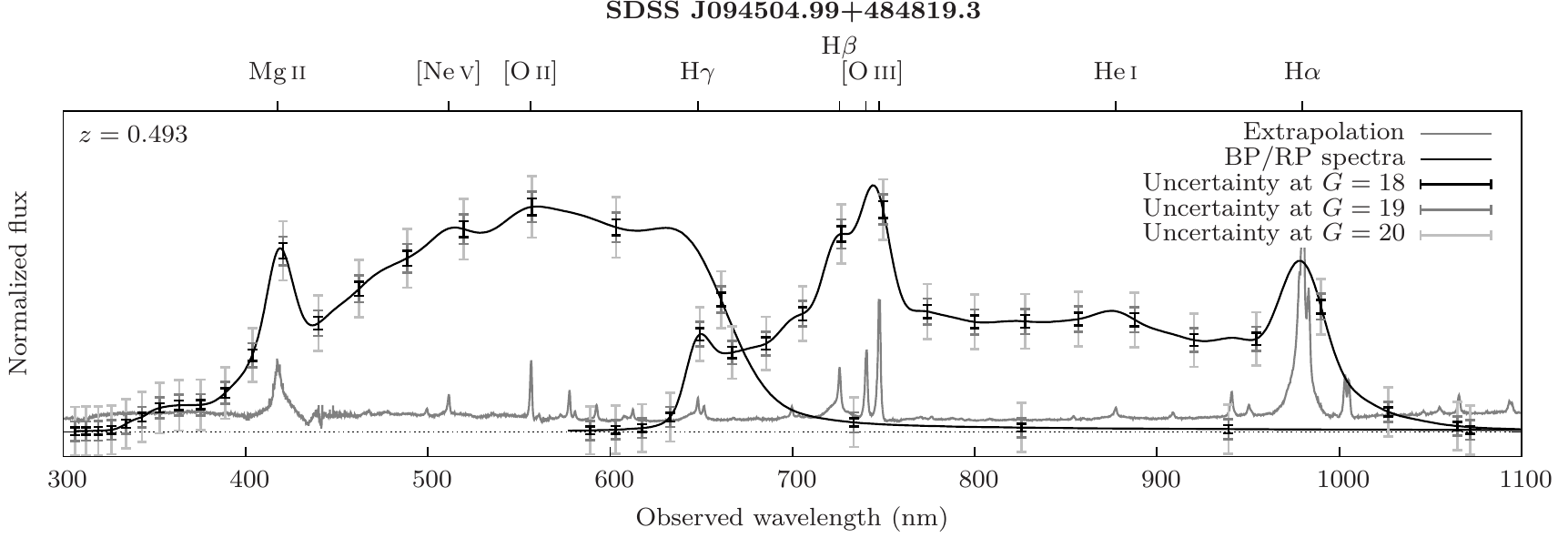}
\caption{Example of produced BP/RP spectrum and associated flux uncertainties for $G$ magnitudes of $18$, $19$ and $20$. Spectra normalization allows to fairly compare the noise levels that are present amongst the various provided magnitudes. We can straightly figure out the effect of the global instrumental response and of the LSF convolution applied to the extrapolated spectrum as respectively producing characteristic bell shapes as well as very broadened emission lines within the resulting BP/RP spectra. The noticed non-uniform spectral resolution arises from the varying wavelength dispersion function of each of the BP/RP spectrophotometers.}
\label{fig:bprp_spectra}
\end{figure*}

	Being now able to model the entire instrumental response along with the associated uncertainties, we first choose to normalize our input spectra to $G$ magnitudes where we expect QSOs to be observed, that is at $G = \lbrace 18, 18.5, 19, 19.5, 20 \rbrace$. This normalization allows us to study the behaviour of the implemented methods under an increasing level of noise. BP/RP spectra were then produced based on the most up-to-date instrument model coming from tests carried out by EADS Astrium (later renamed Airbus Defence and Space) during the commissioning phase of the satellite. We have to note that this instrument model still has to be updated in order to match the actual operational condition of the satellite even if the latter is not expected to vary too much from the model we used. Noisy spectra can then be obtained by adding the appropriate random Gaussian noise (i.e. having a variance of $\sigma_i^2$) to each of the (noise-free) spectral fluxes, $s_i$. An example of produced BP/RP spectra is illustrated in figure \ref{fig:bprp_spectra}.

\section{Astrophysical parameter determination}
\label{sec:ap_determination}

\begin{figure*}
\includegraphics{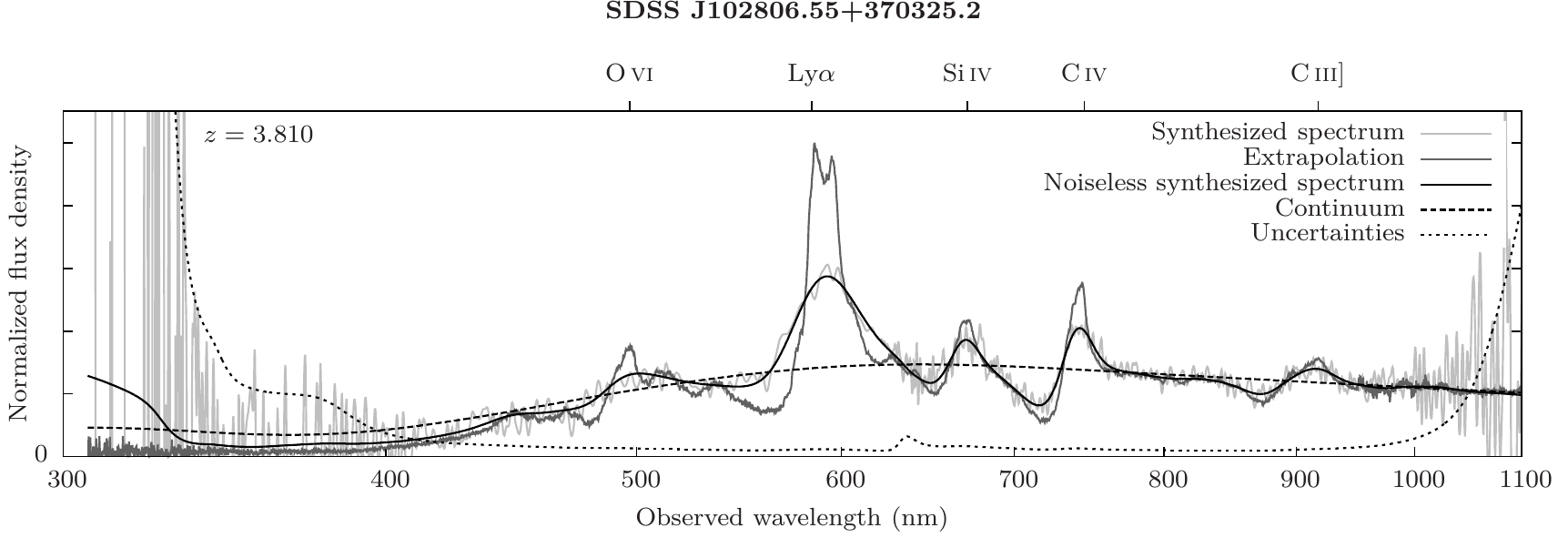}
\caption{Illustrative synthesized spectrum coming from the linear combination of the interpolated and flat-fielded BP/RP spectra of magnitude $G = 19$. The observed BAL features allows to highlight the capability of pre-processing we used in order to recover the general shape of the extrapolated spectrum. Flux uncertainties were also pre-processed accordingly, the BP/RP transition region being unequivocally recognized as a bump around $635$ nm. We have to note that the noiseless synthesized spectrum comes from a similar pre-processing applied to a noise-free version of the BP/RP spectra.}
\label{fig:bprp_synthesized}
\end{figure*}

	The bell shape of the BP/RP spectra prevents us from using both algorithms described in section \ref{sec:methods}. This is even more damageable given the fact that these are not sampled on a logarithmic wavelength scale and that the wavelength coverage of each pixel is not uniform. In order to tackle these problems, a resampling of the spectra fluxes and uncertainties was first performed using cubic spline interpolation \citep{press2002}. The uniform logarithmic sampling we used, $\Delta \log_{10} \lambda = 1.75 \times 10^{-4}$, ensuring a sampling on the redshift that is better than $0.003$, which is comparable to human expertise, while producing a reasonable amount of $7.7 \times 10^3$ sampled points in the final templates (assuming that $z \leq 6$). Note that such a logarithmic interpolation will obviously introduce covariances between the resulting samples. Nevertheless, given that the specific resampling we used stands in a wavelength range where the corresponding logarithmic function is approximately linear and that both the number of samples within the BP/RP spectra and within the synthesized spectra are of the same order of magnitude, we will have that these covariances will be restricted to close neighbouring samples while having moderate magnitudes. These will consequently have a limited impact on the resulting predictions and will hence be ignored in the following. The division of these interpolated spectra by a flat BP/RP spectrum (i.e. coming from a flat input SED) concurrently fixes the bell-shape issue, that is mostly due to the global instrumental response, as well as the problem of the non-uniform wavelength coverage of the pixels, that is due to the inconstant wavelength dispersion function of the BP/RP spectrophotometers. Accordingly, these flat-fielded spectra can then be considered as being approximately proportional to the convolution of the input SED by the LSF over a linear wavelength scale, plus noise. Now, we will have that the resulting spectra will be disjoint although they are overlapping, which would yield to a tremendous loss of efficiency if these were to be considered individually. A more interesting solution stands in the linear combination of the flat-fielded BP/RP spectra according to a given weighting scheme such as to produce a single synthesized spectrum. In more details, if we consider $s_\lambda^\mathrm{bp}$ and $s_\lambda^\mathrm{rp}$ as being the interpolated fluxes of the BP and RP spectra; $\sigma_\lambda^\mathrm{bp}$, $\sigma_\lambda^\mathrm{rp}$, as their associated uncertainties; $F_\lambda^\mathrm{bp}$, $F_\lambda^\mathrm{rp}$, as their corresponding flat BP/RP fluxes and $w_\lambda^\mathrm{bp}$, $w_\lambda^\mathrm{rp}$ as the weighting coefficient used to join these spectra, then we have that the synthesized fluxes, $f_\lambda$, can be represented as
\begin{equation}
\label{eq:bprp_flux_synthesized}
f_\lambda = w_\lambda^\mathrm{bp} \frac{s_\lambda^\mathrm{bp}}{F_\lambda^\mathrm{bp}} + w_\lambda^\mathrm{rp} \frac{s_\lambda^\mathrm{rp}}{F_\lambda^\mathrm{rp}},
\end{equation}
and their associated uncertainties, $\sigma_\lambda$, as
\begin{equation}
\label{eq:bprp_uncertainty_synthesized}
\sigma_\lambda = \left[\left(w_\lambda^\mathrm{bp} \frac{\sigma_\lambda^\mathrm{bp}}{F_\lambda^\mathrm{bp}}\right)^2 + \left(w_\lambda^\mathrm{rp} \frac{\sigma_\lambda^\mathrm{rp}}{F_\lambda^\mathrm{rp}}\right)^2\right]^{\frac{1}{2}}.
\end{equation}
In the context of the present study, the weighting coefficients we selected are given by
\begin{equation}
\label{eq:bprp_weighting_scheme}
w_\lambda^\mathrm{rp} = 1 - w_\lambda^\mathrm{bp} = \transition{620}{650}{\lambda},
\end{equation}
where
\begin{equation}
\label{eq:tanh_transition_weights}
\transition{\lambda_0}{\lambda_1}{\lambda} = \frac{1}{2} \tanh \left[2 \pi \left(\frac{\lambda-\lambda_0}{\lambda_1-\lambda_0}-\frac{1}{2}\right)\right]+\frac{1}{2}
\end{equation}
is the hyperbolic tangent transition function from $\lambda_0$ to $\lambda_1$. The specific weighting used in equation \ref{eq:bprp_weighting_scheme} ensures a smooth transition between the flat-fielded BP and RP spectra while keeping most of their significant regions. A continuum spectrum was then gathered and subsequently subtracted from each synthesized spectrum using a procedure similar to the one described in section \ref{subsec:spectra_extrapolation}. An illustrative synthesized spectrum produced through equations \ref{eq:bprp_flux_synthesized} and \ref{eq:bprp_uncertainty_synthesized} is shown in figure \ref{fig:bprp_synthesized}.

	Having now exploitable input spectra, we decided to split our input spectral library into two parts, the first part being used as a `learning set' (LS1) in order to produce the PCA templates upon which we will base the analysis of the second part, the `test set' (TS1). Conversely, the second part will be subsequently used as a learning set (LS2) for the analysis of the first part (TS2). This two-fold cross validation procedure will finally provide the APs for the whole set of observations as if these were gathered based upon a totally independent dataset. The splitting criterion we used relies on the uniform selection of half the input library sorted according to the QSOs redshifts such as to ensure an even repartition of the latter amongst these two data sets.
	
\begin{figure*}
\includegraphics{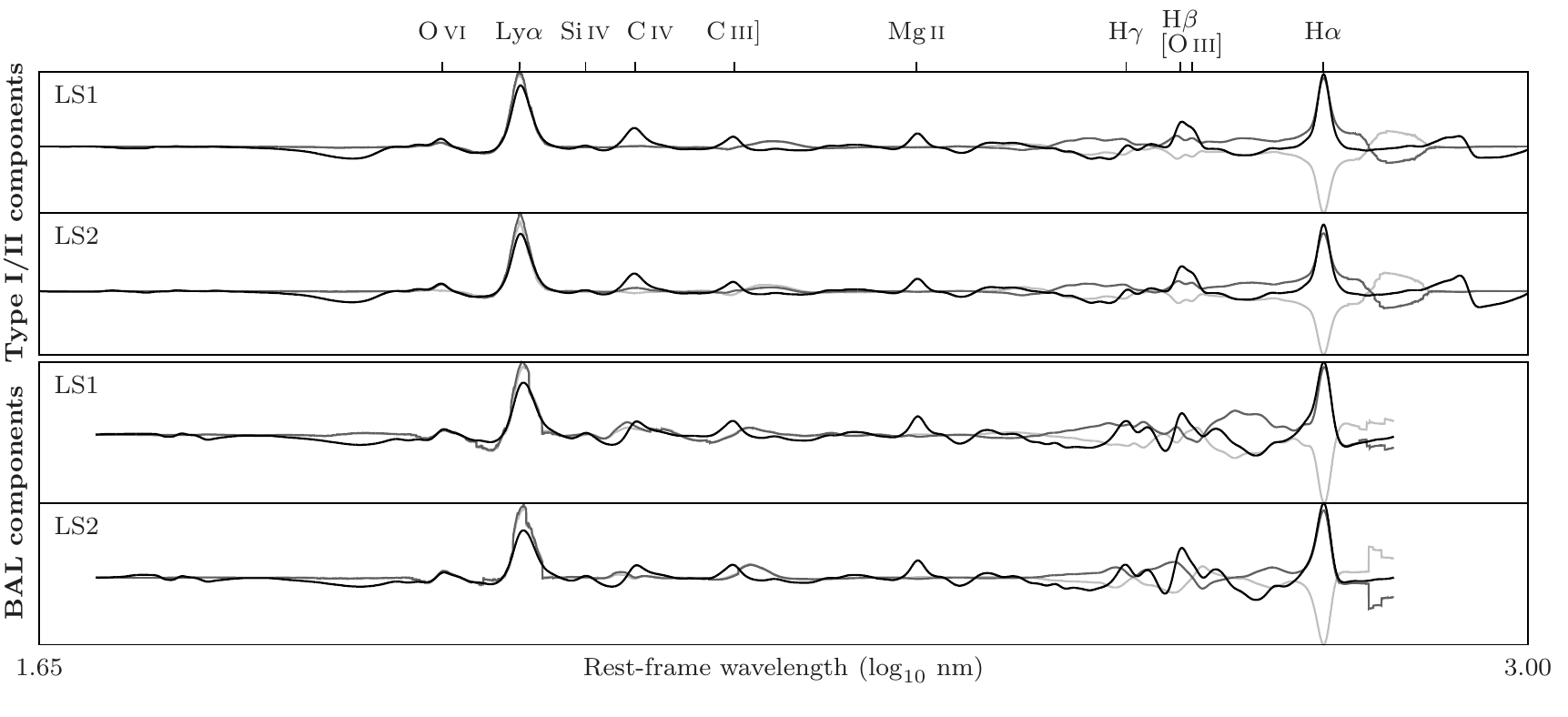}
\caption{Mean observations (black lines) and first two principal components coming from the PCA decompositions of BP/RP synthesized spectra out of LS1 and LS2 regarding type I/II QSOs (up) and BAL QSOs (down). Both learning sets show extremely good agreement for both QSO types, the noticed discrepancies in the BAL components at wavelength longer than the H$\gamma$ emission line being explained by the small number of BAL QSOs having $z \leq 1.42$ ($\sim 1\%$ of the learning sets). We can further notice that most of the BAL features standing between the \ion{O}{vi} and \ion{C}{iv} emission lines were retained from the DR12Q library (see figure \ref{fig:pca_extrapolation}) while the removal of the fine QSO structures by the BP/RP instrumental convolution now allows the first BAL templates to model the entire emission lines.}
\label{fig:bprp_synthesized_pca}
\end{figure*}
	
	Two sets of rest-frame PCA templates were then produced for each of the learning set according to the QSO type (type I/II or BAL). These were based on the noise-free; continuum-subtracted and synthesized BP/RP spectra having both un-normalized $G$ magnitudes and SNR $> 1$ within the extrapolation procedure. From the noiseless nature of these input spectra, we had to select custom weights associated with the synthesized fluxes, $f_\lambda$, as
\begin{equation}
\label{eq:bprp_pca_templates_weights}
w_\lambda = \transition{330}{380}{\lambda} \transition{1050}{925}{\lambda} \left[0.7 \left(2 \transition{620}{650}{\lambda} - 1 \right)^2 + 0.3 \right],
\end{equation}
where the first two terms practically reflect the limited confidence we set on the spectra edges due, for example, to the potential inaccuracies in the spectra extrapolation or to low fluxes within the flat BP/RP spectra leading to numerical instabilities and where the last term stands for the uncertainties introduced in the BP/RP spectra combination. Figure \ref{fig:bprp_synthesized_pca} provides the mean observation and first two principal components of the synthesized spectra of the type I/II and BAL QSOs for both learning sets.

	It is worth to mention that, at first glance, it might seems misleading from the point of view of the validation process to retrieve PCA components from synthesized spectra which are themselves based on the linear combination of templates. Nevertheless, let us first remind that it is one of our assumptions that any (noiseless) DR12Q quasar spectrum can be fairly represented as the linear combination of a sufficient number of such templates. Secondly, we have to note that BP/RP spectra come from the instrumental convolution of the extrapolated spectra in the observed wavelengths. The latter being then set on rest frame, we will have that the resulting PCA components will have to reflect the averaged convolution applied over the whole observed wavelengths. 	Finally, this convolution will have the effect of smoothing the high-frequency components from the extrapolated spectra. These being concurrently the main source of unexplained variance within the DR12Q templates, we expect the produced library to be consistent regarding an hypothetical real noise-free BP/RP spectral library.
	
\begin{table}
\caption{Warning flags used in the redshift selection procedure. These can be combined through bitwise OR operator.}
\label{tbl:zwarning}
\begin{tabular}{lcl}
\hline
Warning flag & Value & Condition(s) for rising \\
\hline
\verb+Z_AMBIGUOUS+ & 1 & More than one peak have both \\
& & $0 < z < 6$ and $\chi_r^2(z) > 0.85$\\
\verb+Z_LOWCHI2R+ & 2 & $\chi_r^2(z) < 0.9$ \\
\verb+Z_LOWZSCORE+ & 4 & $\mtxt{Z}{score}(z) < 0.9$ \\
\verb+Z_NOTOPTIMAL+ & 8 & We did not choose the optimal\\
& & peak (i.e. $\chi_r^2(z) < 1$) \\
\hline
\end{tabular}
\end{table}

	The extracted PCA components were then used in order to produce their CCF against the noisy synthesized BP/RP spectra of magnitude $G = \lbrace 18, 18.5, 19, 19.5, 20 \rbrace$ through the algorithm described in section \ref{subsec:wpcorr}. The redshift identification being based on the CCF peak having a corresponding redshift in the range $0 < z < 6$; a $\chi_r^2(z) > 0.85$ and a minimal scaled distance from the ideal point $(1,1) \in (\chi_r^2(z), \mtxt{Z}{score}(z))$ as given by
\begin{equation}
\label{eq:scaled_distance_chi2r_zscore}
d(z) = \sqrt{\left(0.8 \left[1-\chi_r^2(z)\right]\right)^2 + \left(0.2 \left[1-\mtxt{Z}{score}(z)\right]\right)^2}.
\end{equation}
The selected redshift was then flagged for potential inaccuracies in the peak selection according to the values provided within Table \ref{tbl:zwarning}. Constants used in the peak selection procedure as well as within Table \ref{tbl:zwarning} are purely empirical and based on a visual inspection procedure.

	The optimal number of PCA components to use was chosen as a trade-off between the ratio of explained variance; the ability of the templates to model BAL QSOs and the potential overfit of the observations coming from the use of a too large number of templates. This overfitting being characterized by frequent ambiguities in the corresponding CCFs that eventually results in a large number of erroneous redshift predictions (though these will have a non-zero warning flag). Tests performed on each learning set show that the use of $3$ PCA components is a satisfactory compromise between these constraints that ultimately leads to a ratio of explained variance of $94.6\%$ (LS1) and $93.42\%$ (LS2) regarding the type I/II QSOs and of $86.22\%$ (LS1) and $88.14\%$ (LS2) regarding the BAL QSOs.

	The BAL QSO identification is based on the comparison of the value of the CCF peak we selected using type I/II templates, $y(z)$, against the value of the CCF peak selected from BAL templates, $\mtxt{y}{b}(\mtxt{z}{b})$, through
\begin{equation}
\label{eq:bal_discriminant}
\mtxt{p}{b} = \frac{\mtxt{y}{b}(\mtxt{z}{b})}{\mtxt{y}{b}(\mtxt{z}{b})+y(z)},
\end{equation}
where $z$ is the redshift selected from type I/II templates and $\mtxt{z}{b}$ is the redshift selected from BAL templates. Though straight classification between these two types of QSOs is sometimes practical, it is however commonly motivated by the specific needs of the end-users. As an example, studying the physics of BAL QSOs will require an extremely pure subset of observations (e.g. with $\mtxt{p}{b} > 0.7$) while a re-observation survey can easily deals with a `hint' on the BAL nature of the observed QSOs (e.g. with $\mtxt{p}{b} > 0.5$). Still, the frequent discrepancies observed between the redshift predicted based upon these two kinds of templates enforce us to use such a classification. Accordingly we will consider, in the following, $\mtxt{z}{b}$ as the default reshift whenever $\mtxt{p}{b} > 0.55$ and $G \leq 19$ while keeping $\mtxt{p}{b}$ as a discriminant value for further application-specific classification. The effect of the thresholding of this discriminant value on the resulting ratio of correctly/incorrectly classified observations will be deferred to section \ref{subsec:bal_binary_classification}.

\begin{figure}
\includegraphics{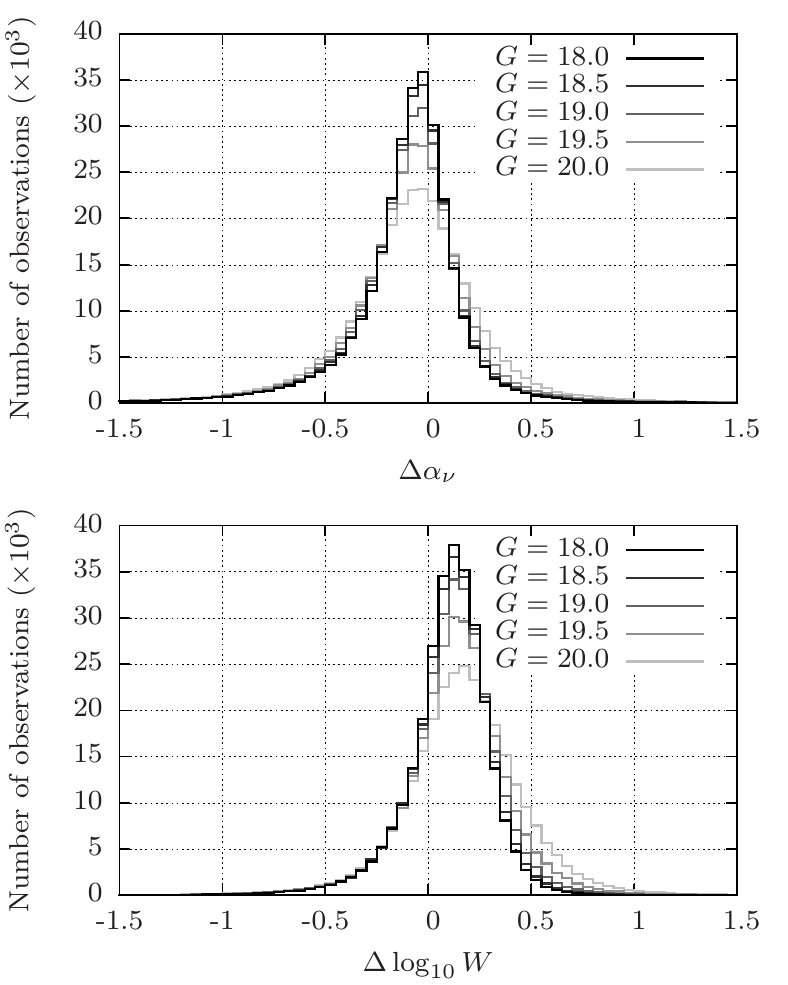}
\caption{Histograms of the differences between the continuum slopes, $\alpha_\nu$, and total emission line equivalent widths, $W$, predicted based upon DR12Q spectra and synthesized spectra regarding various normalization magnitudes.}
\label{fig:alpha_nu_ew}
\end{figure}


	The slope of the QSO continuum corresponds to the spectral index, $\alpha_\nu$, as defined by
\begin{equation}
\label{eq:alpha_nu}
f_\lambda \propto \lambda^{-\alpha_\nu-2}
\end{equation}
or more compactly expressed in terms of frequency, $\nu$, as $f_\nu \propto \nu^{\alpha_\nu}$. This index is obtained from the fit of a power law function to the observations over wavelength regions that are commonly devoid from emission/absorption features, that is: $145$--$148$; $170$--$180$; $200$--$260$; $325$--$470$ and $525$--$625$ nm. The exact procedure employs a $k$-sigma clipping algorithm (with $k = 3$, $\sigma = 1$) such as to underweigh iron emission blends as well as other fortuitous absorption/emission structures by a factor $100$. This procedure was applied to both the input DR12Q spectra and to the synthesized BP/RP spectra as a way to fairly compare the resulting predictions while discarding any bias that can be due to the differences in the used algorithms. Because of their high numerical complexities, non-linear optimization algorithms were not used for the least-squares solution of equation \ref{eq:alpha_nu}. Rather, each power-law function was fitted through a linear regression of the wavelengths against the fluxes by taking the logarithm of both sides of the latter equation. Although this choice seems to be harmless from the point of view of the DR12Q spectra, synthesized spectra will have to cope with the large amount of discarded samples coming from the frequent negative fluxes encountered within the spectra edges. These discarded samples leading to a large bias towards positive fluxes (see figure \ref{fig:bprp_synthesized}, for example), we consequently decided to reject samples standing outside the observed region $350$--$950$ nm for these specific spectra.

	Finally, the total equivalent width of the emission lines can be represented as 
\begin{equation}
W = \int \frac{e_\lambda}{c_\lambda} d\lambda
\label{eq:total_el_ew}
\end{equation}
where $c_\lambda$ is the continuum slope we fitted based upon equation \ref{eq:alpha_nu} and $e_\lambda$ are the emission lines fluxes, the latter being set to $f_\lambda - c_\lambda$ if $\lambda$ belongs to an emission line region and to zero otherwise. The identification of these emission line regions is based on parts of the spectrum where smoothed fluxes coming from a $45$ points wide Savitsky-Golay filtering \citep{press2002} stands higher than the continuum. In agreement with  what was previously done, equation \ref{eq:total_el_ew} was integrated over the interval $380$--$925$ nm regarding DR12Q spectra and over the interval $350$--$950$ nm regarding the synthesized spectra. The rest-frame total equivalent width being hence straightly given by $\mtxt{W}{rest} = W / (z+1)$.

\begin{table}
\caption{Mean predicted continuum slope, $\alpha_\nu$, and total emission line equivalent width, $W$, based on DR12Q spectra and synthesized spectra with various magnitude $G$.}
\begin{tabular}{cccc}
\hline
& & Mean $\alpha_\nu$ & Mean $\log_{10} W$  \\
\hline
\multicolumn{2}{l}{DR12Q spectra} & $-0.697 \pm 0.626$ & $1.74 \pm 0.236$  \\
\multicolumn{2}{l}{Synthesized spectra} & & \\
& $G = 18.0$ & $-0.561 \pm 0.721$ & $1.668 \pm 0.298$ \\
& $G = 18.5$ & $-0.560 \pm 0.722$ & $1.662 \pm 0.306$ \\
& $G = 19.0$ & $-0.561 \pm 0.724$ & $1.648 \pm 0.321$ \\
& $G = 19.5$ & $-0.569 \pm 0.726$ & $1.626 \pm 0.344$ \\
& $G = 20.0$ & $-0.584 \pm 0.723$ & $1.590 \pm 0.377$ \\
\hline
\end{tabular}
\label{tbl:alpha_nu_ew_res}
\end{table}

\begin{figure*}
\includegraphics{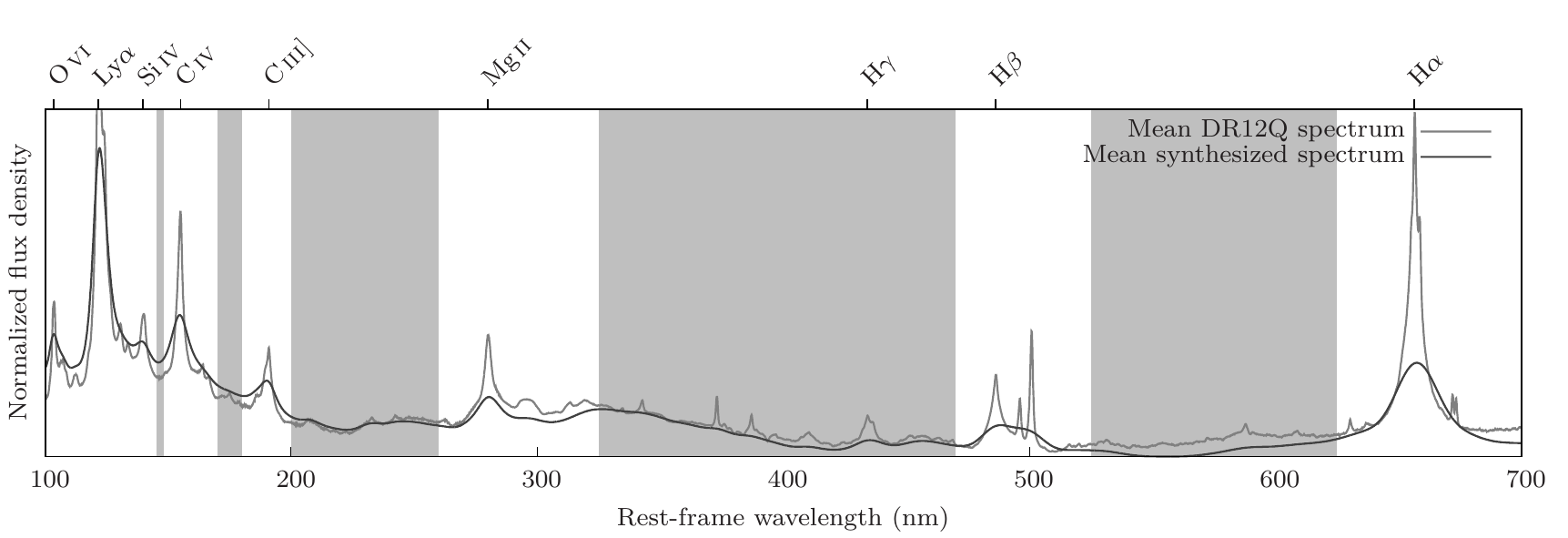}
\caption{Mean spectra coming from the aggregation of DR12Q spectra normalized such a to have a weighted norm equal to one and from synthesized spectra being normalized in the same way. Flux densities were normalized between $200$ and $260$ nm. Shaded regions correspond to wavelength ranges that were used in order to fit the QSO continuum slopes through equation \ref{eq:alpha_nu}. One can notice the higher flux densities of the synthesized mean spectrum within the continuum regions $145$--$148$ and $170$--$180$ nm when compared to the DR12Q mean spectrum while the lower flux encountered within regions $325$--$470$ and $525$--$625$ nm often yield a red bias during the continuum fitting procedure.}
\label{fig:alpha_nu_bias}
\end{figure*}

	The results of these continuum slope and total emission line equivalent width determination procedures are summarized within Table \ref{tbl:alpha_nu_ew_res} and figure \ref{fig:alpha_nu_ew} regarding DR12Q spectra and synthesized spectra of various normalizing magnitudes. We can easily see that the continuum slopes predicted from synthesized spectra tend to be bluer than those of the DR12Q spectra. This bias comes from the spread of the \ion{Si}{iv} and \ion{C}{iv} emission lines over the continuum region $145$--$148$ nm and, to a lesser extent, from a similar spread of the \ion{C}{iv} and \ion{C}{iii]} emission lines over the continuum region $170$--$180$ nm as depicted within figure \ref{fig:alpha_nu_bias}. The noticed relation between the flattening of the continuum slope and the normalizing magnitude comes from the increasing number of negative samples that are rejected within the continuum regions $325$--$470$ nm and $525$--$625$ nm where faint fluxes are usually found and which ultimately tend to artificially redden those regions. Also, we can observe that the total equivalent widths of the emission lines predicted based upon synthesized spectra are underestimated compared to the ones predicted using DR12Q spectra. The reason for this similarly stands within the globally overestimated continuum flux as well as from the reddening of the spectrum at wavelengths longer than $300$ nm according to the magnitude. Note that the miss of some narrow emission lines because of the LSF convolution also tends to lessen the predicted $W$. The reader should hence pay a careful attention to these systematic effects once using these measurements.
	
	Given these shortcomings, one might rightfully wonder whether the use of non-linear optimization algorithms worths to be envisaged in order to predict the continuum slope of the QSOs at the expense of a ten times longer execution time. Doing so  will provide us with a mean value of the continuum slopes of $-0.691 \pm 0.657$ for the DR12Q spectra and a correlation factor of $0.966$ if these are compared to the results of our approach. In the case of synthesized spectra, these numbers become respectively $-0.563 \pm 0.721$, $-0.562 \pm 0.737$, $-0.564 \pm 0.737$, $-0.575 \pm 0.739$ and $-0.601 \pm 0.738$ for the mean continuum slopes of magnitudes $G = \left\lbrace 18, 18.5, 19, 19.5, 20\right\rbrace$ with associated correlation factors of $0.988$ for $G \leq 19$ and of $0.985$, $0.975$ for $G = 19.5$ and $20$, respectively. The observed flattening of the predicted continuum slopes at $G > 19$ ironically comes from the non-rejection of the negative fluxes from the red part of the spectra which tends to give larger weights to these regions (i.e. the fraction of red fluxes being then more significant). While this effect will have a negligible (but still noticeable) impact on the $G \leq 19$ predictions because of the sufficient SNR of the red part of the spectra at these magnitudes (e.g. the fit of the red part of the spectra providing a good approximation of the continuum slopes at these magnitudes), it will have a deleterious impact on fainter magnitudes where the red part of the spectra is often better approximated by a flat curve. This effect gets further amplified through the subsequent rejection of the blue fluxes by the $k$--sigma clipping algorithm. Let us still mention that this non-linear approach remains the most rigorous in a statistical point of view though the strong similarities noticed in both approaches and their common difficulties in predicting the continuum slopes of faint sources do not justify its use regarding its larger time consumption.

\section{Performance comparison}
\label{sec:comparison}

	The performances of our approach were assessed in comparison with the Extremely Randomized Trees learning method \citep[hereafter ERT]{geurts2006}. While classical tree-based learning methods usually try to find, at each node, a splitting criterion (i.e. an attribute and a threshold within this attribute) that is such that the learning set of observations associated with this node is split at best with respect to a given score measure (e.g. variance reduction in regression problem or information gain in classification problem), the ERT instead picks up $K$ random attributes as well as a random threshold associated with each of these attributes in order to select the one maximizing the provided score measure. This procedure is then recursively repeated until the number of learning set observations in all leaf nodes falls under a given limit, $\mtxt{n}{min}$. The averaged prediction of a set of $N$ trees then allows to subsequently lessen the variance of the model (i.e. the sensitivity of each individual tree to the used learning set). The choice of this specific method mainly comes from both its fast learning phase as well as from its high performances regarding other competing methods like Artificial Neural Networks or Support Vector Machine while having only a few numbers of parameters to tune. Let us also note that this method is the one that is presently in use within the QSOC software module in order to predict most of the QSO APs \citep{bailerjones2013}.
	
	First of all, let us mention that the QSO continuum slope and that the total equivalent width of the emission lines will not be considered within this performance comparison because these can be straightly predicted based on observable quantities. Regarding the adjustment of the parameters of the ERT models, tests have shown that the prediction of the QSO redshift and type are rather insensitive to the $K$ and $\mtxt{n}{min}$ parameters if these stand within reasonable ranges of values. Consequently and according to \citet{geurts2006}, the default values of $K = \mtxt{N}{attr}$, $\mtxt{n}{min} = 5$ and $K = \sqrt{\mtxt{N}{attr}}$, $\mtxt{n}{min} = 2$ were accepted respectively for the redshift regression problem and BAL classification problem, $\mtxt{N}{attr}$ being here the number of points contained within our BP/RP spectra (i.e. $\mtxt{N}{attr} = 960$ if $\mtxt{N}{over} = 8$). The number of trees to build, $N$, should be ideally as large as possible. Nevertheless, based on time and memory constraints we have, the latter was set to $N = 1000$. 
	
	In the following, the ERT models will be built based upon the noisy learning sets LS1 and LS2 where the observations having a SNR $> 1$ are selected and normalized such as to have a unit norm for the whole set of magnitudes. Their predictions being then gathered from the associated test sets of corresponding magnitude within TS1 and TS2. We have to note that because of selection effects and observational bias within the DR12Q catalogue, neither LS1 nor LS2 will follow a realistic distribution of the redshift \citep{paris2017}. Similarly, these will neither contain a genuine fraction of BAL QSOs \citep{reichard2003,knigge2008,gibson2009}. Consequently, the ERT models that will be built based on these learning sets will be particularly suited for the prediction of the QSO redshifts and type that are the most frequently encountered within LS1 and LS2. In that sense, these will constitute data-oriented models whose predictions on TS1 and TS2 will be optimistic when compared to those based on real observations. Finally, we can further note that the weighted phase correlation algorithm (hereafter WPC) is not sensitive to this data unbalancing and that the associated results will remain valid irrespectively of the actual APs distribution we will encounter.
	
\begin{figure*}
\includegraphics{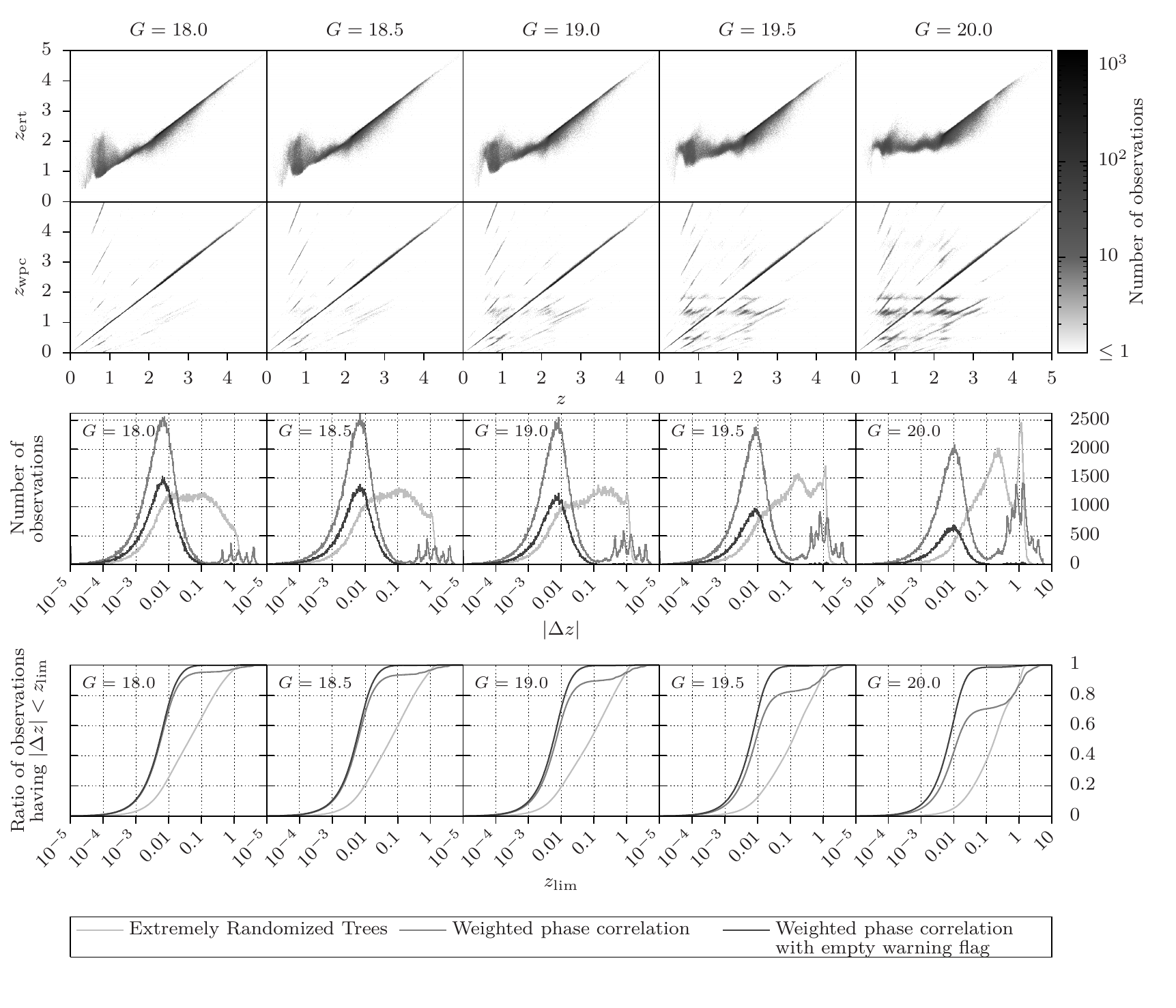}
\caption{(top) Distribution of the predicted redshifts coming from ERT models, $\mtxt{z}{ert}$, and WPC model, $\mtxt{z}{wpc}$, for quasars with $G$ magnitudes equal to $18$, $18.5$, $19$, $19.5$ and $20$ with respect to the DR12Q redshift, $z$. (middle) Histogram of the absolute error, $|\Delta z|$, between DR12Q redshifts and redshift predicted based upon ERT models, WPC models and WPC models while having an empty warning flag (see table \ref{tbl:zwarning}). (bottom) Ratio of observations having an absolute error on the redshift that is lower than some threshold limit as given by $\mtxt{z}{lim}$.}
\label{fig:redshift_result}
\end{figure*}

\subsection{Redshift determination}
\label{subsec:redshift_determination}

	The distribution of the predicted redshifts against DR12Q redshifts is given within the upper part of figure \ref{fig:redshift_result} for the case of the ERT predictions as well as for the case of the WPC predictions regarding the various normalizing magnitudes. We can already notice a trend of the ERT predictions standing at $z \leq 2$ to be driven towards $\mtxt{z}{ert} \approx 2.3$ where stand most of our learning set observations. This effect is particularly noticeable at $z \approx 0.8$ where stands our second most numerous source of QSOs and it further tends to strengthen along with an increasing magnitude (though these misclassified observations will typically have strong associated uncertainties). To a lesser extent, we may also note an opposite trend where the observations standing between $z = 2$ and $z = 3$ tend to be underestimated. These effects potentially reflect the inability of our models to fully grab the information that is present within our learning sets and/or the incompleteness of the latter. The fact that very high redshift objects ($z > 3$) get correctly predicted presumably comes from the entrance of the Ly$\alpha$ forest within the observed wavelength range where the extremely faint fluxes found therein allow to unequivocally characterize these observations.
	
	Although roughly performed here, the analysis of the results coming from machine learning methods often suffer from a lack of physical significance and interpretation that mostly arises because of their underlying complexity. Furthermore, these methods strongly depend on the completeness of the learning set we used in order to build them. In an illustrative purpose, let us consider that a given QSO spectrum is getting a correct redshift prediction from such a model, suppose now that a similar spectrum has a slightly higher redshift which results in a mean shift by a few pixels in the observed spectrum, then nothing ensures us that this shifted spectrum will get a correct prediction from the previous model since this ultimately depends on whether or not a somewhat similar spectrum was encountered within the used learning set. According to this, a learning method dedicated to the redshift determination of QSOs should ideally be based on a learning set of observations covering the vast majority of QSO shapes and characteristics over the entire range of redshift we are looking for. With these arguments in mind, we may still suppose that the ERT models we used (and in a broader sense, any model based on machine learning method) are not the best suited in predicting the redshift of QSOs.

	Regarding the redshift distribution from WPC (see figure \ref{fig:redshift_result}, up), we can readily notice a tighter dispersion of the errors when compared to the ERT predictions with median absolute errors of $0.0057$, $0.0061$, $0.0069$, $0.0088$ and $0.0130$ for the WPC predictions of magnitudes $G = \lbrace 18,18.5,19,19.5,20 \rbrace$, respectively, and corresponding ERT median absolute errors of $0.0419$, $0.0586$, $0.0755$, $0.1164$ and $0.1723$ (see figure \ref{fig:redshift_result}, middle). Similarly, $4.83\%$, $6.66\%$, $10.56\%$, $17.6\%$ and $28.93\%$ of the observations have a catastrophic prediction on their redshift (i.e. $|\Delta z| > 0.1$) within the WPC for the same set of magnitudes while the corresponding ratios of ERT observations become respectively $35.04\%$, $40.37\%$, $45.09\%$, $52.97\%$ and $63.19\%$ (see figure \ref{fig:redshift_result}, down).

\begin{table}
\caption{Ratio of observations triggering warning flags and associated ratio of observations having $\abs{\Delta z} > 0.1$ while triggering these warning flags.}
\label{tbl:zwarning_dz}
\begin{tabular}{cccccc}
\hline
\multicolumn{6}{c}{\bfseries Ratio of observations triggering warning flags}\\
Triggered & \multicolumn{5}{c}{$G$ magnitude}\\
warning flag & $18.00$ & $18.50$ & $19.00$ & $19.50$ & $20.00$\\
\hline
\texttt{Z\_AMBIGUOUS}& $0.1125$& $0.1393$& $0.1853$& $0.2882$& $0.3935$\\
\texttt{Z\_LOWCHI2R}& $0.0084$& $0.0107$& $0.0149$& $0.0310$& $0.0440$\\
\texttt{Z\_LOWZSCORE}& $0.4198$& $0.4602$& $0.5153$& $0.5833$& $0.6733$\\
\texttt{Z\_NOTOPTIMAL}& $0.0330$& $0.0427$& $0.0594$& $0.1115$& $0.1636$\\
Any& $0.4620$& $0.5099$& $0.5772$& $0.6709$& $0.7758$\\
None& $0.5380$& $0.4901$& $0.4228$& $0.3291$& $0.2242$\\
\hline
\multicolumn{6}{c}{\bfseries Ratio of observations having $\abs{\Delta z} > 0.1$}\\
Triggered & \multicolumn{5}{c}{$G$ magnitude}\\
warning flag & $18.00$ & $18.50$ & $19.00$ & $19.50$ & $20.00$\\
\hline
\texttt{Z\_AMBIGUOUS}& $0.2934$& $0.3221$& $0.3644$& $0.4290$& $0.5071$\\
\texttt{Z\_LOWCHI2R}& $0.5607$& $0.5744$& $0.5795$& $0.6085$& $0.6474$\\
\texttt{Z\_LOWZSCORE}& $0.0964$& $0.1240$& $0.1814$& $0.2684$& $0.3884$\\
\texttt{Z\_NOTOPTIMAL}& $0.4434$& $0.4690$& $0.5032$& $0.5478$& $0.6077$\\
Any& $0.1018$& $0.1279$& $0.1798$& $0.2592$& $0.3688$\\
None& $0.0023$& $0.0028$& $0.0043$& $0.0064$& $0.0142$\\
\hline
\end{tabular}

\end{table}

	Most of the WPC errors come from mismatches between emission lines. These mainly consist in the confusions of H$\beta$ with \ion{C}{iii]}; \ion{Mg}{ii} with Ly$\alpha$; \ion{Mg}{ii} with \ion{C}{iv}; \ion{Mg}{ii} with \ion{C}{iii]} and \ion{C}{iv} with Ly$\alpha$. We have to note that these mismatches do not constitute by themselves real cases of degeneracy but rather arise because of the effect of noise on the emission lines identification as we will soon see. Still, this effect is already unambiguously depicted in figure \ref{fig:redshift_result} (top), where the number of observations suffering from such an emission line mismatch problem tends to increase along with an increasing magnitude. In the same figure, we may also note that low SNR spectra ($G \geq 19$) tend to produce constant predictions at $\mtxt{z}{wpc} \approx 0.5$, $1.35$ and $1.8$. These correspond to the fit of deviant fluxes from spectra edges by the H$\alpha$, \ion{C}{iv} and Ly$\alpha$ emission lines, respectively. Though unavoidable, most of these errors will come along with a non-empty warning flag (see Table \ref{tbl:zwarning_dz}) which offers the possibility to discard these insecure predictions. By doing so, we rejected $46.2\%$, $50.99\%$, $57.72\%$, $67.09\%$ and $77.58\%$ of the total number of observations regarding magnitudes $G = \lbrace 18,18.5,19,19.5,20 \rbrace$, respectively. This leads to corresponding median absolute errors of $0.0053$, $0.0055$, $0.0058$, $0.0063$ and $0.0072$ and associated ratios of catastrophic redshift predictions of $0.23\%$, $0.28\%$, $0.43\%$, $0.64\%$ and $1.42\%$ for the same set of magnitudes.

	From our previous discussion, we can notice that the performances we gained were achieved at the expense of a very high rejection rate of the observations having a non-empty warning flag. The distribution of these warning flags amongst the observations is given in Table \ref{tbl:zwarning_dz} along with their associated ratio of catastrophic redshift prediction once triggered. We first have to note that \texttt{Z\_LOWZSCORE} is the most frequently triggered warning flag amongst these observations and is hence the one that contributes at most in their removal. The reason for this stands in the fact that the $\mtxt{Z}{score}$ measure is primarily designed such as to be sensitive to the presence of all the emission lines that are theoretically covered at a given redshift estimate. The miss of one such a line can be attributed either to the wrong redshift estimate we made or to its misidentification owing to its noise or its strong damping by the instrumental convolution, though the right redshift was selected. This misleading distinction is clearly depicted in Table \ref{tbl:zwarning_dz} where solely $9.64\%$ of the observations having $G = 18$ and \texttt{Z\_LOWZSCORE} flag set comes along with $\abs{\Delta z} > 0.1$, thus arguing for the frequent misidentification of some emission lines while at $G = 20$ this ratio becomes $38.84\%$, hence consisting in a larger fraction of effective redshift confusions. Secondly, the \texttt{Z\_AMBIGUOUS} flag is frequently set because of the intrinsic degeneracy existing in the prediction of the redshift of quasars albeit we can assess from Table \ref{tbl:zwarning_dz} that in $70.66\%$ of the cases the right redshift is selected amongst these ambiguous solutions at $G = 18$ while this ratio drops to $49.29\%$ at $G = 20$. For completeness, we have to mention that $49.29\%$ of successful identifications is still better than the ratio that would be obtained from a random selection of the solution given that the observations having \texttt{Z\_AMBIGUOUS} flag set often consist in more than two ambiguous solutions (see figure \ref{fig:redshift_result}, for example). We can further note that, once an ambiguity is detected (i.e. \texttt{Z\_AMBIGUOUS} warning flag triggered), the optimal peak of the CCF is commonly selected as the most probable redshift estimate as these do not additionally trigger a \texttt{Z\_NOTOPTIMAL} warning flag. Now, if a sub-optimal peak of the CCF is selected, then $55.66\%$ of the observations comes along with $\abs{\Delta z} \leq 0.1$ at $G=18$ while this ratio becomes $39.23\%$ at $G = 20$. Given that the latter observations must have a $\mtxt{Z}{score}$ that is greater or equal to the one associated with the optimal peak of the CCF in order to be selected, these eventually reveal the effective degeneracy that exists in the redshift prediction of quasars once based on low SNR spectra. Finally, the \texttt{Z\_LOWCHI2R} warning flag is rarely triggered given the strong constraints we set on it (see equation \ref{eq:scaled_distance_chi2r_zscore}). This decision is further supported by the fact that the associated ratio of catastrophic redshift prediction stands to be the highest amongst the whole set of warning flags.
	
\begin{figure*}
\includegraphics{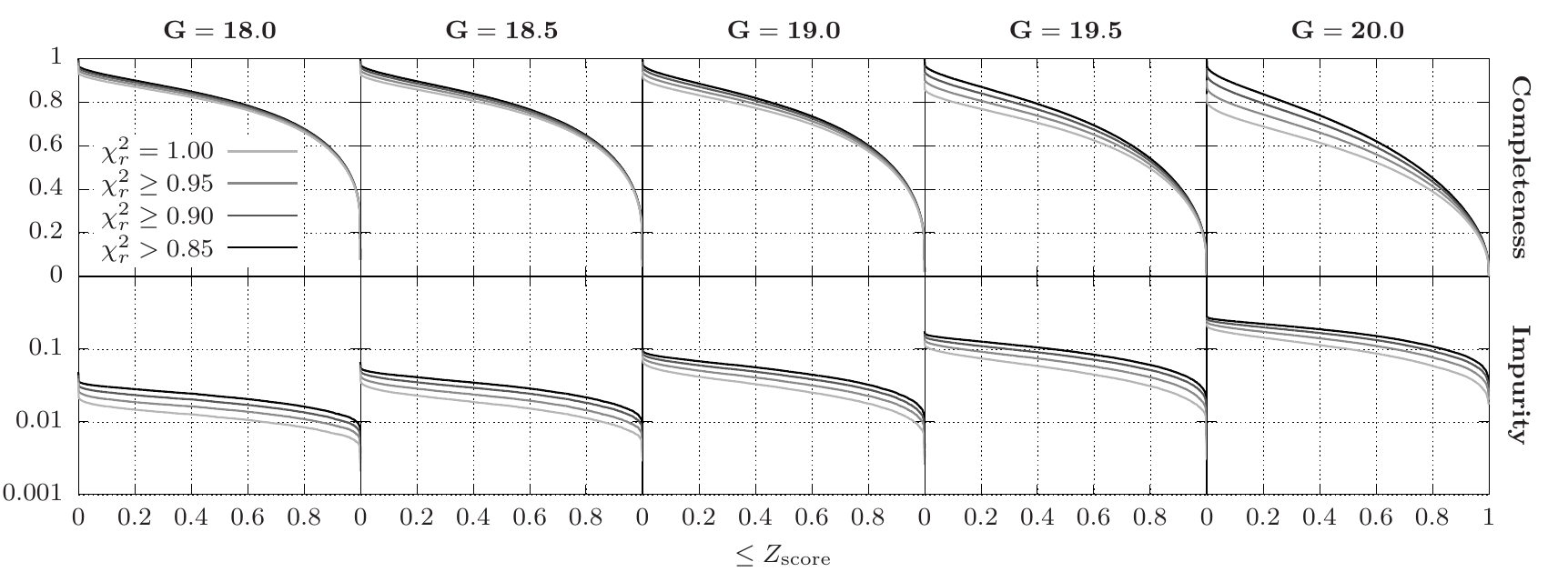}
\caption{Ratio of observations having $\mtxt{Z}{score}$ and $\chi_r^2$ greater or equal to the provided values (i.e. completeness) and ratio of observation having $\abs{\Delta z} > 0.1$ amongst the latter (i.e. impurity).}
\label{fig:completeness_imputity}
\end{figure*}
	
	As pointed out within Section \ref{sec:ap_determination}, the thresholds that were used in order to trigger the \texttt{Z\_LOWZSCORE} and \texttt{Z\_LOWCHI2R} warning flags are somewhat arbitrary and other values might be better suited regarding the specific needs of the end user. This is particularly true given that these were shown to have a strong impact on the trade-off between the completeness and the impurity of our predictions as we have just seen. These ratios of completeness and impurity being given in figure \ref{fig:completeness_imputity} for varying thresholds on $\chi_r^2$ and $\mtxt{Z}{score}$. Note that we do not considered the \texttt{Z\_AMBIGUOUS} and \texttt{Z\_NOTOPTIMAL} warning flags in this analysis given that $\chi_r^2 < 1$ automatically implies that both these flags are set. Also remind that we required $\chi_r^2 > 0.85$ such as to limit the number of ambiguous solutions that are potentially associated with each observation.
	
\subsection{BAL binary classification}
\label{subsec:bal_binary_classification}

\begin{figure*}
\includegraphics{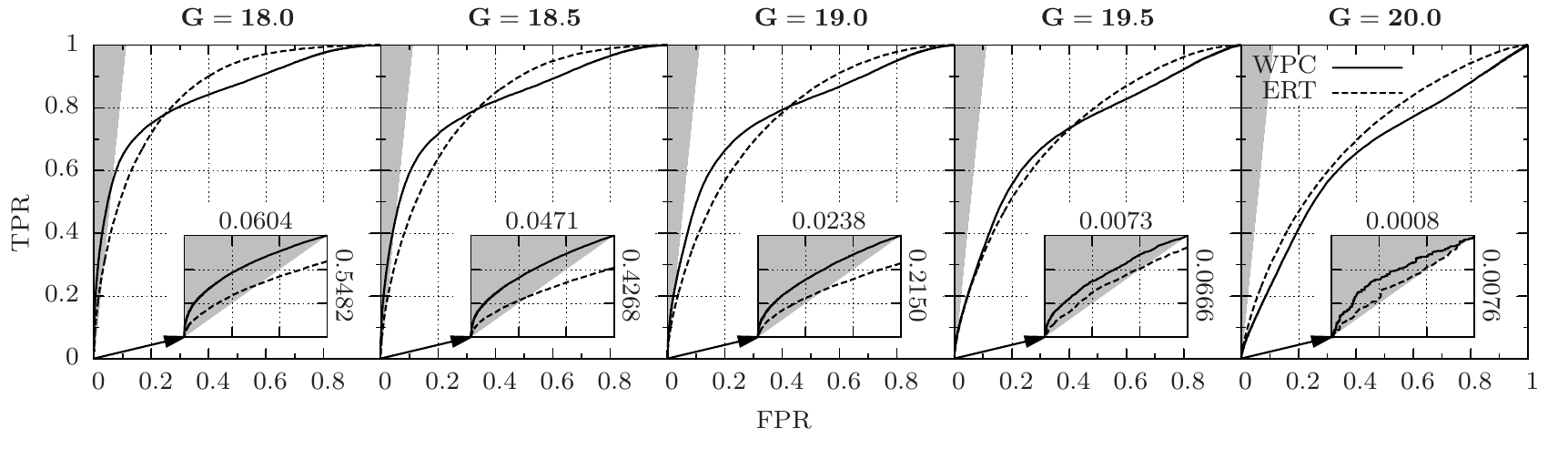}
\caption{ROC curves of BAL binary classification regarding the WPC and ERT models for quasars with $G$ magnitudes equal to $18$, $18.5$, $19$, $19.5$ and $20$. Shaded areas correspond to regions were the accuracy stands higher or equal to the one coming from a constant type I/II classifier. The points of the ROC curves associated with the optimal accuracies stand at the highest distances from the border of these regions while being contained in it.}
\label{fig:bal_result}
\end{figure*}

	The data unbalancing has a particularly insidious impact on the analysis of the results coming from the binary classification of BAL QSOs. Indeed, based on the fact that solely $9.95\%$ of the DR12Q observations are BAL QSOs, a model that will systemically classify the observations as type I/II QSO would then provide a satisfactory ratio of correctly classified observations (i.e. an \textit{accuracy}) of $90.05\%$ while no BAL QSO will be identified. Consequently, this ratio will not constitute an objective analysis tool if considered alone. We will hence use two additional and complementary statistical measures that were specifically designed for the analysis of the performance of binary classifiers. First, the \textit{true positive rate} (hereafter TPR) will here denotes the fraction of BAL QSOs that are correctly identified by a given model. It constitutes an estimator of the probability of detection of the BAL QSOs by this model. Secondly, the \textit{false positive rate} (hereafter FPR) will denote the fraction of type I/II QSOs that are wrongly classified as BAL QSOs. A perfect binary classifier should hence have TPR $= 1$ along with FPR $= 0$. Note that both these statistical measures can be adjusted by varying the user-defined threshold that was set either on $\mtxt{p}{b}$, for the case of the WPC, or on the number of trees that voted for the BAL class regarding the ERT. By doing so and reporting the corresponding TPR against FPR we obtain the so-called \textit{Receiver Operating Characteristics} (ROC) curve as depicted within figure \ref{fig:bal_result} for the case of the WPC and ERT models for quasars with magnitudes $G = \lbrace 18,18.5,19,19.5,20 \rbrace$. These curves allow to straightly compare the performances of these two competing models while depending neither on the data unbalancing, nor on the specific thresholds we used. The area under the ROC curve being then often taken as a fair indicator of their global performances.
	
	Now, like many data reduction pipelines, our primary objective will be to optimize the accuracy of our model with respect to the fraction of BAL QSOs that will be encountered amongst the real observations. We will then have to take into account the potential unbalancing that will be present within the \textit{Gaia} observations. However, because of the uncertainties surrounding the selection effect from DSC as well as the observational bias, this unbalancing is not known \textit{a priori}. Consequently, we decided to consider a fraction of BAL QSOs, $\mtxt{r}{b}$, equal to the one that is present within the DR12Q catalogue (i.e. $\mtxt{r}{b} = 0.0995$). The presented accuracies should hence be updated once a realistic ratio will be available though the general conclusions drawn out of these are not supposed to change (assuming that $\mtxt{r}{b}$ remains small). We can then easily figure out that the regions of the ROC curves where the accuracy is constant correspond to lines whose equations are given by
\begin{equation}
\mathrm{TPR} = \frac{1 - \mtxt{r}{b}}{\mtxt{r}{b}} \times \mathrm{FPR} + C.
\label{eq:caccuracy}
\end{equation}
Our goal will then be to find the point(s) of the ROC curve that intersect such a line while maximizing $C$. Note that the trivial case where $C = 0$ corresponds to the accuracy that would be obtained by a constant type I/II classifier which is thereby always achievable. Stated otherwise, the point(s) of the ROC curve having an optimal associated accuracy correspond(s) to the one (or those) whose distance to the line of constant type I/II accuracy is the greatest while being on its left side.

\begin{table}
\caption{Optimal accuracies of the ERT and WPC models for quasars with $G$ magnitudes of $18$, $18.5$, $19$, $19.5$ and $20$ along with their associated thresholds, TPR and FPR.}
\label{tbl:optimal_accuracy}
\begin{tabular}{rcccc}
\hline
\multicolumn{5}{c}{\textbf{ERT}} \\
$G$ magnitude & Threshold & TPR & FPR & Accuracy \\
\hline
$18.00$ & $552$ & $0.15553$ & $0.01048$ & $90.644\%$ \\
$18.50$ & $604$ & $0.07920$ & $0.00448$ & $90.428\%$ \\
$19.00$ & $604$ & $0.05204$ & $0.00345$ & $90.252\%$ \\
$19.50$ & $598$ & $0.02857$ & $0.00235$ & $90.119\%$ \\
$20.00$ & $617$ & $0.00352$ & $0.00029$ & $90.058\%$ \\
\hline
\multicolumn{5}{c}{\textbf{WPC}} \\
$G$ magnitude & Threshold & TPR & FPR & Accuracy \\
\hline
$18.00$ & $0.5341$ & $0.31992$ & $0.01666$ & $91.725\%$ \\
$18.50$ & $0.5517$ & $0.21295$ & $0.01105$ & $91.169\%$ \\
$19.00$ & $0.5843$ & $0.09140$ & $0.00437$ & $90.562\%$ \\
$19.50$ & $0.6102$ & $0.03348$ & $0.00203$ & $90.198\%$ \\
$20.00$ & $0.6486$ & $0.00413$ & $0.00023$ & $90.069\%$ \\
\hline
\end{tabular}
\end{table}

\begin{figure}
\includegraphics{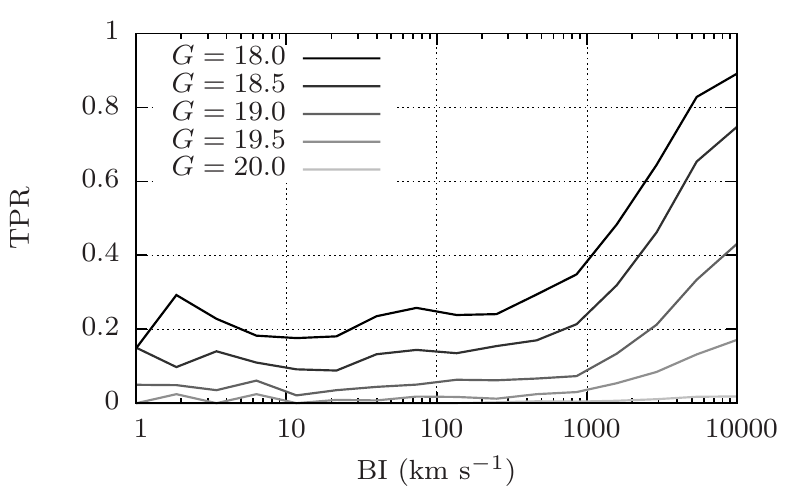}
\caption{Comparison of the Balnicity index of the \ion{C}{iv} trough with respect to the TPR of the WPC models regarding BAL QSOs with various magnitudes.}
\label{fig:bi_tpr}
\end{figure}

	Figure \ref{fig:bal_result} focuses on the regions of the ROC curves where the accuracy stands higher than the one of a constant type I/II classifier. From our previous discussion, we can readily see that the WPC models have overall better accuracies when compared to the ERT models for the whole set of $G$ magnitudes. The best achievable accuracies being summarized in Table \ref{tbl:optimal_accuracy} along with their associated thresholds, FPR and TPR for $G$ magnitudes of $18$, $18.5$, $19$, $19.5$ and $20$. The extremely low TPR found therein can be explained by both the relatively low SNR of the synthesized spectra as well as by the removal of most of the narrow absorption features by the LSF convolution and/or by the under-sampling of these spectra. The effect of noise can be readily recognized based on figure \ref{fig:bal_result} where the ROC curves tend to match the ones that would be obtained from a  random classifier (i.e. a diagonal line) as we increase the $G$ magnitude. This translates as a drop of the point of optimal accuracy along the ROC curves which consists in both a lower TPR and a compensating lower FPR (see Table \ref{tbl:optimal_accuracy}). In extreme cases, BAL QSOs become barely identifiable with a probability of detection within the WPC of $3.348\%$ for $G$ magnitude of $19.5$ and of $0.413\%$ for the case of $G = 20$. Figure \ref{fig:bi_tpr} compares the TPR of the WPC with the Balnicity index of the \ion{C}{iv} trough \citep[hereafter BI]{weymann1991} for the various normalizing magnitudes. This BI can be seen as a modified equivalent width of the BAL absorption occurring in the blue part of the \ion{C}{iv} emission line.  We can notice a strong dependence of the TPR according to BI, which reflects the difficulty in identifying BAL QSOs having narrow absorption features. We can finally notice that if one can afford to have a high FPR, then the ERT provides a better TPR than the WPC. This would be the case, for example, if we would like to filter the \textit{Gaia} catalogue by keeping most of the the BAL QSOs while rejecting a still significant number of type I/II QSOs.

\section{Discussion}
\label{sec:discussion}

	Although already fully operational, the presented software module may still experience some minor improvements that will be summarized in the remainder. First, we did not consider any extinction by the interstellar medium. The associated correction relies on the availability of a wavelength-dependent extinction law such as the one of \citet{fitzpatrick1999} as well as on a map of galactic extinction like the one that will be produced by the Total Galactic Extinction software module from CU8 \citep{bailerjones2013}. The total equivalent width of the emission lines as well as the continuum slope from equation \ref{eq:alpha_nu} might benefit from this correction. Nevertheless, due to the fact that the continuum slopes we subtracted from our synthesized spectra are purely empirical (see section \ref{sec:ap_determination}), these will also contain most of the encountered extinction. Accordingly, this correction is not expected to bring any major improvement on the prediction of the redshift of QSOs nor on the subsequent calculation of the BAL discriminant value, $\mtxt{p}{b}$. Furthermore, based on the fact that most of the DR12Q spectra stand at relatively high galactic latitude (i.e. $|b| > 30\degr$) where the extinction is weak, the spectra we used in this study were not much affected by this extinction. A more challenging objective would be to enable the prediction of this extinction based on the BP/RP spectra of quasars. This problem is currently being investigated but seems to be hardly attainable because of the degeneracy existing between the extinction curve and the intrinsic continuum slope of the QSOs. Secondly, the computation of a $\chi^2$ value from the optimal point of the CCF (see equation \ref{eq:wpcorr_chi2z_bis}) can straightly allow to send feedback about the potential misclassification of the quasars we received from the DSC module.

\section{Conclusion}
\label{sec:conclusion}

	We have described in the present work the processing of the BP/RP spectra coming from the \textit{Gaia} satellite in order to determine the astrophysical parameters of quasars within the QSOC module of the CU8 coordination unit from the DPAC. These astrophysical parameters encompass: the redshift of the QSOs, their continuum slopes, the total equivalent width of their emission lines and whether or not these are broad absorption lines (BAL) QSOs. We have highlighted the necessity to have fast and reliable algorithms such as to deal with the huge amount of spectra that \textit{Gaia} will provide as well as with their limited signal-to-noise ratio and resolution. We have introduced two already developed algorithms, namely the weighted principal component analysis and the weighted phase correlation, that were specifically designed in order to fulfil both these mentioned objectives and whose combination allows to securely predict both the redshift of the QSOs and to set a discriminant on their type. We have presented the construction of a semi-empirical library of BP/RP spectra based on the \textit{Gaia} instrumental convolution of the observations coming from the Sloan Digital Sky Survey which were extrapolated in order to cover the wavelength range of the BP and RP spectra. We saw the pre-processing that is required in order for these BP/RP spectra to be fully exploitable by our algorithms as well as the methods we used for predicting the various astrophysical parameters. Some systematic bias were noticed within the prediction of the continuum slopes and of the total equivalent width of the emission lines. These bias can be mostly explained by both the spread of the \ion{Si}{iv}, \ion{C}{iv} and \ion{C}{iii]} emission lines over the continuum regions situated between $145$--$148$ nm and $170$--$180$nm as well as by the rejection of the negative fluxes that are usually found within the red part of the pre-processed spectra.
	
	A comparison with the currently used machine learning method showed that our approach is the one of predilection for the determination of the redshift of the quasars while benefiting from a straight physical significance as well as from strong diagnostic tools on the potential errors that may arise during predictions. Cross validation tests showed that $95.17\%$, $93.34\%$, $89.44\%$, $82.4\%$ and $71.07\%$ of the observations come along with an absolute error on the predicted redshift that is lower than $0.1$ for the case of quasars with $G$ magnitudes equal to $G = \lbrace 18, 18.5, 19, 19.5, 20\rbrace$. These ratios become respectively $99.77\%$, $99.72\%$, $99.57\%$, $99.36\%$ and $98.580\%$ once the insecure predictions are discarded based on the triggering of some warning flags. We explored the repartition of these warning flags amongst the observations and studied the effect of setting customized warning thresholds on the trade-off between the completeness and the impurity of our predictions. Our methods were proved to yield the best ratio of correctly classified observations regarding the identification of BAL QSOs assuming that these will be observed much less frequently than the type I/II QSOs. Machine learning methods may still provide a better probability of detection of these BAL QSOs at the expense of much higher contamination rates. Finally, we have that $91.725\%$, $91.1069\%$, $90.562\%$, $90.198\%$ and $90.069\%$ of the observations were correctly classified by our methods regarding quasars with $G$ magnitudes of $18$, $18.5$, $19$, $19.5$ and $20$, respectively.

\section*{Acknowledgements}
The author acknowledges support from the ESA PRODEX Programme `Gaia-DPAC QSOs' and from the Belgian Federal Science Policy Office.

Funding for SDSS-III has been provided by the Alfred P. Sloan Foundation, the Participating Institutions, the National Science Foundation and the U.S. Department of Energy Office of Science. The SDSS-III web site is http://www.sdss3.org/.

SDSS-III is managed by the Astrophysical Research Consortium for the Participating Institutions of the SDSS-III Collaboration including the University of Arizona, the Brazilian Participation Group, Brookhaven National Laboratory, Carnegie Mellon University, University of Florida, the French Participation Group, the German Participation Group, Harvard University, the Instituto de Astrofisica de Canarias, the Michigan State/Notre Dame/JINA Participation Group, Johns Hopkins University, Lawrence Berkeley National Laboratory, Max Planck Institute for Astrophysics, Max Planck Institute for Extraterrestrial Physics, New Mexico State University, New York University, Ohio State University, Pennsylvania State University, University of Portsmouth, Princeton University, the Spanish Participation Group, University of Tokyo, University of Utah, Vanderbilt University, University of Virginia, University of Washington and Yale University. 

\bibliographystyle{mnras}
\bibliography{paper}

\bsp
\label{lastpage}
\end{document}